\RequirePackage{fix-cm}
\documentclass[smallcondensed]{svjour3}       % onecolumn (second format)
\smartqed  % flush right qed marks, e.g. at end of proof
\usepackage{graphicx}
\usepackage{color}

\usepackage{setspace}

\usepackage{exscale,lscape}
\usepackage[english]{babel}
\usepackage{rawfonts}
\usepackage{latexsym}
\usepackage[cp850]{inputenc}
\usepackage{epsfig}
\usepackage[intlimits]{amsmath}
\usepackage{slashbox}
\usepackage{multirow}

\newcommand{\bol}[1]{\mbox{\boldmath$#1$}}

\newcommand{\bSigma}{\bol{\Sigma}}

\newcommand{\bmu}{\bol{\mu}}
\newcommand{\tbm}{\tilde{\bol{\mu}}}

\newcommand{\bx}{\mathbf{X}}
\newcommand{\bQ}{\mathbf{Q}}

\newcommand{\by}{\mathbf{Y}}

\newcommand{\bL}{\mathbf{L}}

\newcommand{\bR}{\mathbf{R}}

\newcommand{\bz}{\mathbf{z}}

\newcommand{\bw}{\mathbf{w}}

\newcommand{\bii}{\mathbf{1}}
\newcommand{\bzero}{\mathbf{0}}
\newcommand{\bI}{\mathbf{I}}

\newcommand{\Var}{\mbox{Var}}

\newcommand{\tbx}{\tilde{\bx}}
\newcommand{\bA}{\bol{A}}
\newcommand{\brx}{\breve{\bx}}
\newcommand{\brm}{\breve{\bol{\mu}}}
\newcommand{\brA}{\breve{\bA}}

\newcommand{\tbF}{\tilde{\mathbf{\Phi}}}
\newcommand{\bF}{\mathbf{\Phi}}
\newcommand{\bn}{\boldsymbol{\nu}}

\newcommand{\eps}{\pmb{\varepsilon}}

\begin{document}

\title{A Closed-Form Solution of the Multi-Period Portfolio Choice Problem for a Quadratic Utility Function}

\titlerunning{Closed-Form Solution of Multi-Period Portfolio Choice for Quadratic Utility}        % if too long for running head

\author{Taras Bodnar        \and
        Nestor Parolya \and %etc.
				Wolfgang Schmid
}

\authorrunning{Taras Bodnar, Nestor Parolya and Wolfgang Schmid} % if too long for running head

\institute{T. Bodnar \at
							Institut f\"{u}r Mathematik\\
              Humboldt Universit\"{a}t zu Berlin \\
              \email{bodnar@math.hu-berlin.de}           %  \\
           \and
           N. Parolya \at
              Institute of Empirical Economics\\
              Leibniz University of Hannover\\
              Hannover\\
              \email{nestor.parolya@ewifo.uni-hannover.de}
              \and
           W. Schmid \at
           Department of Statistics\\
              European University Viadrina\\
              Frankfurt (Oder)\\
              \email{schmid@europa-uni.de}
}

\date{Received: date / Accepted: date}

\maketitle

\begin{abstract}
In the present paper, we derive a closed-form solution of the multi-period portfolio choice problem for a quadratic utility function with and without a riskless asset.  All results are derived under weak conditions on the asset returns. No assumption on the correlation structure between different time points is needed and no assumption on the distribution is imposed. All expressions are presented in terms of the conditional mean vectors and the conditional covariance matrices.

If the multivariate process of the asset returns is independent it is shown that in the case without a riskless asset the solution is presented as a sequence of optimal portfolio weights obtained by solving the single-period Markowitz optimization problem. The process dynamics are included only in the shape parameter of the utility function. If a riskless asset is present then the multi-period optimal portfolio weights are proportional to the single-period solutions multiplied by time-varying constants which are depending on the process dynamics. Remarkably, in the case of a portfolio selection with the tangency portfolio the multi-period solution coincides with the sequence of the simple-period solutions. Finally, we compare the suggested strategies with existing multi-period portfolio allocation methods for real data.
\end{abstract}
\newpage
\section{Introduction}

\noindent Nowadays, the portfolio selection problem plays an important role in financial research. A number of papers are devoted to questions like, e.g., how an optimal portfolio can be constructed, monitored, and/or estimated by using historical data (see, e.g., Alexander and Baptista (2004) , Golosnoy and Schmid (2007), Bodnar (2009)), what is the influence of parameter uncertainty on the portfolio performance (cf., Okhrin and Schmid (2006) , Bodnar and Schmid (2008)), how do the asset returns influence the portfolio choice (see, e.g., Jondeau and Rockinger (2006), Menc\'{i}a and Sentana (2009), Adcock (2009), Harvey et al. (2010), Amenguala and Sentana (2010)), how is it possible to estimate the characteristics of the distribution of the asset returns (see, e.g., Jorion (1986), Wang (2005), Frahm and Memmel (2010)), how can the structure of optimal portfolio be statistically justified (Gibbons et al. (1989), Britten-Jones (1999), Bodnar and Schmid (2009)).

In a seminal paper from 1952, H. Markowitz presented the idea of an optimal portfolio selection by taking into account the trade-off between the portfolio expected return and its risk which is measured by the portfolio variance. The idea of Markowitz's approach is to minimize the portfolio variance for a given level of the expected return. This method is equivalent to the so-called mean-variance utility maximization problem. Although the suggested methodology is quite simple, it provides us the most commonly used solution of the single-period (static) portfolio choice problem that remains very popular today (see, e.g. Brandt (2010)).

Although the case of a long-term investment horizon is of greater importance in practice much less has been done in that area. The first formulation of the multi-period portfolio selection problem has already been given in the book of Markowitz (1959) followed by the papers of Mossin (1968), Samuelson (1969), Merton and Samuelson (1974). Although it is heavily discussed in recent literature (see, e.g., Konno et al. (1993), Li and Ng (2000), Steinbach (2001), Leippold, Trojani and Vanini (2004), Brandt and Santa Clara (2006), Edirisinghe and Patterson, (2006), \c{C}elikyurt and \"{O}zekici (2007), \c{C}anako$\breve{\text{g}}$lu and \"{O}zekici, (2009), Kilianov\'{a} and Pflug (2009), Skaf and Boyd (2009)), to the best of our knowledge, a closed-form solution is not available in the general case up to now. Only under the assumption of independence closed-form solutions are presented by Li and Ng (2000) and Leippold et al. (2004). For more general models, the solution is frequently determined by a numerical procedure (see, e.g. Dantzig and Infanger (1993), van Binsbergen and Brandt (2007), Mansini et al. (2007), K\"{o}ksalan and \c{S}akar (2014)). Brandt and Santa Clara (2006) suggested a solution of the multi-period portfolio selection problem assuming that the portfolio weights can be presented as a linear function of certain state variables. This assumption leads to a simplification of the optimization problem. Note that the solution is only a local maximum which could differ from the global one. Finally, the solution of the multi-period portfolio selection problem in continuous time is given by Duffie and Richardson (1991), Yan and Li (2008), A\"{i}t-Sahalia et al. (2009), Basak and Chabakauri (2010), Marzban et al. (2013) among others.

We contribute to the existing literature by deriving the closed-form solution of the dynamic portfolio choice problem with and without a riskless asset under rather weak assumptions. The only conditions imposed on the distributions of the asset returns are the existence of the conditional mean vectors and of the conditional covariance matrices. No assumptions about the correlation structure between different time points or about the distribution of the asset returns, like normality, are needed. The suggested method can be applied for both stationary and non-stationary stochastic models.  The results are obtained assuming that the investor makes his decision on the basis of the quadratic utility function. This is one of the most commonly used procedures since the paper of Tobin (1958) where it is shown that the Bernoulli principle is satisfied for the mean-variance solution only if one of the following two conditions is valid: the asset returns are normally distributed, which is rarely the case in application, or the utility function is quadratic. On the other hand, the quadratic utility function is usually considered as a good approximation of the other utility functions (cf. Brandt et al. (2006)). Moreover, under the additional assumption of independence we show that at each time point the optimal multi-period portfolio weights can be presented in a similar way as the optimal single-period portfolio weights. Both representations differ in the coefficient of risk aversion. Finally, if the asset allocation is based on the tangency portfolio we prove that the solution of the multi-period portfolio selection problem is the same as that obtained by solving the single-period problem at each time point.

In an empirical study we apply the obtained results to real data by comparing the performance of the suggested strategies with existing multi-period portfolio allocation methods. It is shown that the multi-period portfolio strategies based on the approximative solutions perform very well for different values of the coefficient of risk aversion and different investment periods.

The rest of the paper is organized as follows. In Section 2, we introduce the multi-period portfolio choice problem for a quadratic utility function without a riskless asset. The main results of this section are given in Theorem 1 (Section 2) where a closed-form solution of the optimal portfolio weights is given. In Corollary 1 we present the results of Theorem 1 in terms of the classical one-period Markowitz's solution for independent observations. In Section 3, the solution of the multi-period portfolio selection problem for a quadratic utility function with a riskless asset is presented (see Theorem 2). As in the case without a riskless asset the solution for independent returns is proportional to the solution of the corresponding single-period problem at each time point. The process dynamics exclusively influence the proportionality constant (Corollary 2). A very interesting result is obtained for the tangency portfolio. In Theorem 3 we prove for independent returns that the multi-period portfolio selection problem for the tangency portfolio is equivalent to a sequence of the single-period problems. The same solution is obtained in both cases. Because the tangency portfolio is, usually, considered as a market portfolio (see, e.g. Britten-Jones (1999)) in capital asset pricing theory the obtained result is of great importance for practitioners. The findings of the empirical studies are presented in Section 4. The paper concludes in Section 5. All proofs are given in the appendix (Section 6).

\section{Multi-Period Portfolio Choice Problem for a Quadratic Utility Function: Without Riskless Asset}

\noindent In this section we derive a closed-form solution of the multi-period portfolio choice problem with $k$ risky assets for the case of a quadratic utility function. Let
$\bx_t=\left(X_{t,1},X_{t,2},\ldots,X_{t,k}\right)^{\prime}$ denote the vector of the returns of $k$ risky assets and let $E(\bx_t|\mathcal{F}_{t-1})=\bmu_t$ and $Var(\bx_t|\mathcal{F}_{t-1})=\bSigma_t$. $\mathcal{F}_t$ denotes the information set available at time $t$. $\bSigma_t$ is assumed to be a positive definite matrix. Note that for deriving the closed-form solution of the multi-period portfolio selection problem in the present case we need neither any assumption on the correlation structure between different time points nor any distributional assumption. It is only demanded that the conditional covariance matrix of the asset returns exists. The solution exclusively depends on the conditional mean vector and the conditional covariance matrix. These quantities can be calculated depending on the underlying model of the asset return process. For instance, if the investor assumes that the asset returns follow a vector ARMA-GARCH process then the expressions for the conditional mean vector and the conditional covariance matrix can be directly obtained by applying the well-developed theory of the multivariate autoregressive processes and the multivariate conditionally heteroscedastic autoregressive processes (see, e.g., Brockwell and Davis (1991), Engle (1982, 2002), Bollerslev et al. (1988), Bauwens et al. (2006)).

Let $\bw_t=\left(w_{t,1},w_{t,2},\ldots,w_{t,k}\right)^{\prime}$ denote the vector of portfolio weights in period $t$. Note that $\bw^\prime_t\bii=1$ where the vector $\bii$ denotes a $k$-dimensional vector whose components are all equal to 1. Then the wealth of the investor at time $t$ is given by
\begin{equation}\label{wealth}
W_t=W_{t-1}(1+\bw_{t-1}^{\prime}\bx_t)=W_{t-1}\bw^{\prime}_{t-1}\tbx_t \,,
\end{equation}
where $\tbx_{t}=\bii+\bx_{t}$. Then  $E(\tbx_t|\mathcal{F}_{t-1})=\tbm_{t}=\bii+\bmu_t$ and $Var(\bx_t|\mathcal{F}_{t-1})=\bSigma_t$. Later on, we make use of $\tbx_t$ instead of $\bx_t$ for simplifying the presentation of the obtained results. In this section, we deal with an investor who invests his money exclusively into $k$ risky assets and whose investment strategy is based on the quadratic utility function given by
\begin{equation}\label{QUF}
U(W_t)=W_t-\frac{\alpha}{2}W_t^2\,,
\end{equation}
where $\alpha>0$ is the slope parameter of the quadratic utility function. The corresponding relative risk aversion coefficient (RRA) is given by
\begin{equation}\label{gamma}
\gamma_t=\frac{\alpha W_t}{1-\alpha W_t}\,,
\end{equation}
which specifies the attitude of the investor toward risk. Brandt and Santa Clara (2006) considered $\gamma_t$ as a constant obtained by substituting $W_t=1$ in (\ref{gamma}). We follow this procedure in the empirical part of the paper by choosing the value of $\alpha$ in the definition of the utility function (\ref{QUF}) in such a way that $\gamma\in\{5,10,15,20\}$.

The portfolio is allocated at time point $0$ and thereafter reconstructed at time $1,\ldots,T-1$. The planning horizon contains $T$ periods. The investor gets his reward after the final period at time $T$. The aim of the investor is to maximize his final utility of wealth at time point $T$, i.e.
\begin{equation}\label{OP_0}
V(0,W_0,\mathcal{F}_{0})=\max\limits_{\{\bw_s:\bw^{\prime}_s\bii=1\}_{s=0}^{T-1}}E_0[U(W_T)]\,.
\end{equation}
Here $E_t(\cdot)$ stands for the expectation given the information set $\mathcal{F}_t$ available at time $t$.
We assume that short-selling is allowed, i.e. $\bw_i$ could be negative. This problem can be solved recursively.

Let
\begin{equation}\label{tP_t}
V(t,W_t,\mathcal{F}_{t})=\max\limits_{\{\bw_s:\bw^{\prime}_s\bii=1\}_{s=t}^{T-1}}E_t[U(W_T)]\,
\end{equation}
and let $\bw^{*}_{T-t+1}$ denote the optimal portfolio weights at time point $T-t+1$ which is in general a function of optimal weights of the next periods, i.e. of $\bw^{*}_{T-t+2}$, $\bw^{*}_{T-t+3}$, ..., $\bw^{*}_{T-1}$. Following Brandt and Santa-Clara (2006) and Pennacchi (2008) the optimization problem (\ref{OP_0}) can be solved by applying the following Bellman equation at time point $T-t$

\begin{eqnarray}\label{BE}
&&V(T-t,W_{T-t},\mathcal{F}_{T-t})  \nonumber\\
&=&\max\limits_{\bw_{T-t}:\bw_{T-t}^\prime\mathbf{1}=1}E_{T-t}
\Big{[}\max\limits_{\{\bw_s:\bw^{\prime}_s\bii=1\}_{s=T-t+1}^{T-1}}E_{T-t+1}[U(W_T)]\Big{]}\nonumber \\
&=& \max\limits_{\bw_{T-t}:\bw_{T-t}^\prime\mathbf{1}=1}E_{T-t}
\Big{[}V(T-t+1,W_{T-t}\bw^{*\;\prime}_{T-t}\tbx_{T-t+1},\mathcal{F}_{T-t+1})\Big{]}
\end{eqnarray}

for $t=1,\ldots,T$. Note that
\begin{equation}\label{TC}
V(T,W_T,\mathcal{F}_T)=U(W_T)\,.
\end{equation}

First, we derive the closed-form solution for the weights at period $T-1$ and then formulate the general solution in Theorem 1. At time point $T-1$ the application of the terminal condition (\ref{TC}) leads to
\begin{equation}\label{VT}
V(T,W_T,\mathcal{F}_T)=U(W_T)=W_T-\frac{\alpha}{2}W_T^2
\end{equation}
and, hence,
\begin{eqnarray}\label{Vt_1}
&&V(T-1,W_{T-1},\mathcal{F}_{T-1})=\max\limits_{\{\bw_{T-1}:\bw^{\prime}_{T-1}\bii=1\}}E_{T-1}\Big{[}E_T[W_T-\frac{\alpha}{2}W_T^2]\Big{]}\nonumber\\
&=&\max\limits_{\{\bw_{T-1}:\bw^{\prime}_{T-1}\bii=1\}}E_{T-1}\Big{[}W_{T-1}\bw^{\prime}_{T-1}\tbx_{T}
-\frac{\alpha}{2}\left(W_{T-1}\bw^{\prime}_{T-1}\tbx_{T}\right)^2\Big{]}\nonumber\\
&=&\max\limits_{\{\bw_{T-1}:\bw^{\prime}_{T-1}\bii=1\}}\Big{[}W_{T-1}\bw^{\prime}_{T-1}\tbm_{T}
-\frac{\alpha}{2}E_{T-1}\left(W_{T-1}\bw^{\prime}_{T-1}\tbx_{T}\right)^2\Big{]}\nonumber\\
&=&\max\limits_{\{\bw_{T-1}:\bw^{\prime}_{T-1}\bii=1\}}\Big{[}W_{T-1}\bw^{\prime}_{T-1}\tbm_{T}-\frac{\alpha}{2}W^2_{T-1}\bw^{\prime}_{T-1}\bA_{T}\bw_{T-1}
\Big{]}\,,
\end{eqnarray}
where $\bA_{T}=\bSigma_{T}+\tbm_{T}\tbm^{\prime}_{T}$.

The first order conditions for the period $T-1$ are
\begin{equation}\label{FOC}
W_{T-1}\tbm_{T}-\alpha W^2_{T-1}\bA_{T}\bw_{T-1}+\lambda\bii=0 \quad \text{and} \quad \bii^{\prime}\bw_{T-1}=1\,,
\end{equation}
where $\lambda$ is a Lagrange multiplier. Solving (\ref{FOC}) with respect to $\lambda$ and $\bw_{T-1}$ leads to the portfolio weights for period $T-1$ given by
\begin{equation}\label{weights_T-1}
\bw^*_{T-1}=\frac{\bA^{-1}_T\bii}{\bii^{\prime}\bA^{-1}_T\bii}+\frac{1}{\alpha W_{T-1}}\tilde{\bQ}_T\tbm_T \quad \text{with} \quad \tilde{\bQ}_T=\bA^{-1}_T-\frac{\bA^{-1}_T\bii\bii^{\prime}\bA^{-1}_T}{\bii^{\prime}\bA^{-1}_T\bii}\,.
\end{equation}

In Theorem 1, the expressions of the optimal portfolio weights $\bw^*_{T-t}$ at periods 0 to $T-1$ are presented.

\begin{theorem}
Let $\bx_{\tau}=\left(X_{\tau,1},X_{\tau,2},\ldots,X_{\tau,k}\right)^{\prime}$, $\tau=0,\ldots,T$, be a random return vector of $k$ risky assets and let $E(\bx_{\tau}|\mathcal{F}_{\tau-1})=\bmu_{\tau}$ and $Var(\bx_{\tau}|\mathcal{F}_{\tau-1})=\bSigma_{\tau}$ where $\bSigma_{\tau}$ is positive definite. Then for all $t=1,\ldots,T$ the optimal multi-period portfolio weights for the period $T-t$ are given by
\begin{equation}\label{weights_T-t}
\bw^*_{T-t}=\frac{\bA^{-1}_{T-t+1}\bii}{\bii^{\prime}\bA^{-1}_{T-t+1}\bii}+\frac{1}{\alpha W_{T-t}}\tilde{\bQ}_{T-t+1}\tbm_{T-t+1}^*\,,
\end{equation}
with
\begin{equation}\label{tbQ_i}
\tilde{\bQ}_{T-t+1}=\bA^{-1}_{T-t+1}-
\frac{\bA^{-1}_{T-t+1}\bii\bii^{\prime}\bA^{-1}_{T-t+1}}{\bii^{\prime}\bA^{-1}_{T-t+1}\bii}\,,
\end{equation}

\begin{equation}\label{bA}
\bA_{T-t+1}= \left\{
  \begin{array}{l l}
    \bSigma_T+\tbm_T\tbm^\prime_T &\quad\text{for}~~~~t=1\\
    E_{T-t}[V_{T-t+2}\tbx_{T-t+1}\tbx_{T-t+1}^\prime]& \quad \text{for}~~~ t=2,\ldots,T\,,
  \end{array} \right.,
\end{equation}

\begin{equation}\label{mu}
\tbm^*_{T-t+1}= \left\{
  \begin{array}{l l}
    \tbm_T &\quad\text{for}~~~~t=1\\
    E_{T-t}[R_{T-t+2}\tbx_{T-t+1}]& \quad \text{for}~~~ t=2,\ldots,T\\
  \end{array} \right.
\end{equation}
and
\begin{equation}\label{R}
R_{T-t+2}=\frac{\bii^\prime\bA_{T-t+2}^{-1}\tbm_{T-t+2}^*}{\bii^\prime\bA_{T-t+2}^{-1}\bii},~~~V_{T-t+2}=\frac{1}{\bii^\prime\bA_{T-t+2}^{-1}\bii} \quad \text{for}~~~ t=2,\ldots,T\,.
\end{equation}

\end{theorem}

The proof of the theorem is given in the appendix. Theorem 1 shows that the structure of the optimal portfolio weights remains the same during the investment period. In practice, the calculation of the optimal multi-period portfolio weights should be performed by starting with $t=T$ and then proceeding to $t=1$.

It is important to note that if the terminal condition (\ref{TC}) is replaced by
\[V(T,W_T,\mathcal{F}_T)=U(W_T)=W_T-\frac{\alpha}{2}(W_T-E_{T-1}(W_T))^2\]
then its solution is the same as in Theorem 1 except the expression at time point $T-1$. Here, we get
\[\bw^*_{T-1}=\frac{\bSigma^{-1}_T\bii}{\bii^{\prime}\bSigma^{-1}_T\bii}+\frac{1}{\alpha W_{T-1}}\bQ_T\tbm_T \quad \text{with} \quad \tilde{\bQ}_T=\bSigma^{-1}_T-\frac{\bSigma^{-1}_T\bii\bii^{\prime}\bSigma^{-1}_T}{\bii^{\prime}\bSigma^{-1}_T\bii}\,.\]

An important property of the optimal weights is shown in Corollary 1 that is proved in the appendix as well. Here, it is shown that under the assumption of independence the optimal weights have the same structure as the ones that are obtained by solving the single-period portfolio selection problem at each time point with the common quadratic utility function given by
\begin{equation}\label{SPP}
\max \quad \bmu_t^\prime \bw-\frac{\alpha}{2}\bw^\prime \bSigma_t \bw \qquad \text{subject to} \qquad \bw^\prime \bii=1\,.
\end{equation}

In the following we use the notations
$$R_{GMV,i}=\frac{\bii^\prime\bSigma^{-1}_i\bmu_i}{\bii^\prime\bSigma^{-1}_i\bii},~~~~V_{GMV,i}=\frac{1}{\bii^\prime\bSigma^{-1}_i\bii},~~~~\text{and}~~~~s_i=\bmu^\prime_i\bQ_i\bmu_i.$$

\begin{corollary}
Let $\bx_{\tau}=\left(X_{\tau,1},X_{\tau,2},\ldots,X_{\tau,k}\right)^{\prime}$, $\tau=0,\ldots,T$, be a sequence of the independently distributed vectors of $k$ risky assets and let $E(\bx_{\tau})=\bmu_{\tau}$ and $Var(\bx_{\tau})=\bSigma_{\tau}$ where $\bSigma_{\tau}$ is positive definite. Then for all $t=1,\ldots,T$ the optimal multi-period portfolio weights for the period $T-t$ are given by
\begin{equation}\label{weights_T-t_cor}
\bw^*_{T-t}=\frac{\bSigma^{-1}_{T-t+1}\bii}{\bii^{\prime}\bSigma^{-1}_{T-t+1}\bii}+ \alpha_{T-t+1}^{-1}{\bQ}_{T-t+1}\bmu_{T-t+1}\,,
\end{equation}
with
\begin{equation}\label{bQ_i}
\bQ_{T-t+1}=\bSigma^{-1}_{T-t+1}-
\frac{\bSigma^{-1}_{T-t+1}\bii\bii^{\prime}\bSigma^{-1}_{T-t+1}}{\bii^{\prime}\bSigma^{-1}_{T-t+1}\bii}
\end{equation}
and
\begin{eqnarray}\label{alph_T-t+1}
\alpha_{T-t+1}^{-1}&=&\frac{\frac{1}{\alpha W_{T-t}}\left(\prod\limits_{i=T-t+2}^{T}a_i\right)-1-R_{GMV,T-t+1}}{1+s_{T-t+1}}~~\text{with}\\
a_i&=&\frac{1+R_{GMV,i}}{(1+R_{GMV,i})^2+(1+s_i)V_{GMV,i}} \nonumber\,.
\end{eqnarray}
\end{corollary}

The results of Corollary 1 are similar to those of Li and Ng (2000) who solved the multi-period portfolio-selection problem assuming that the asset returns are independent. Corollary 1 states that for solving the multi-period portfolio selection problem with the quadratic utility function (\ref{QUF}) it is enough to solve the single-period problem given in (\ref{SPP}) at each time point $t \in \{0,1,...,T-1\}$ and then to make an adjustment in the expression for the coefficient of the investor's risk aversion $\alpha$. It is very remarkable that the optimal weights at time $T-t$ only depend on the mean vector and the covariance matrix at time $T-t+1$ and the risk aversion $\alpha_{T-t+1}$. The mean vectors and the covariance matrices at time points $\tau>T-t+1$ have an influence on the optimal weights only over the quantity $\alpha_{T-t+1}$.

Because in the case of the single-period portfolio selection problem the solution of (\ref{SPP}) lies on the efficient frontier where $\alpha$ specifies the location of the optimal portfolio within the efficient frontier, the same behavior can be observed in the case of the multi-period portfolio allocation problem. The only difference is that the efficient frontier is time-varying since the mean vectors $\bmu_{\tau}$ and the covariance matrices $\bSigma_{\tau}$ are used for its construction. Then the optimal portfolio is obtained by choosing $\alpha_{\tau}$ as specified in Corollary 1. In the special case when the process $\{\bx_{\tau}\}$ consists of independent observations with time-invariant mean vector and time-invariant covariance matrix the efficient frontier remains the same during the whole investment period. Nevertheless, the optimal portfolio obtained by solving the multi-period portfolio selection problem is time-varying because $\alpha_{\tau}$ is a function of future wealths and it is not constant.

Hence, in the case of investing exclusively into risky assets, the solution of the multi-period portfolio selection problem for the quadratic utility function is not equivalent
to the solution of the corresponding $T$ single-period allocation problems. Even if we assume that $\{\bx_t\}$ consists of independent observations and the mean vector and the covariance matrix are constant during the investment period, the risk aversion is time-varying.

The results of Corollary 1 possess another important application. Using the relationship between the mean-variance utility optimization problem (\ref{SPP}) and the Markowitz optimization problem we get the formulation of the later in the multi-period case. At time point $T-t$ it is given by
\begin{eqnarray}\label{SPP_M}
&&\min \quad \bw^\prime \bSigma_{T-t+1} \bw \qquad \text{subject to} \\
&&\bmu_{T-t+1}^\prime \bw= R_{GMV,T-t+1}+ \frac{\left(\frac{1}{\alpha W_{T-t}}\left(\prod\limits_{i=T-t+2}^{T}a_i\right)-1-R_{GMV,T-t+1}\right)s_{T-t+1}}{1+s_{T-t+1}} \,, \nonumber\\
&&\bw^\prime \bii=1 \nonumber
\end{eqnarray}
for $t =1,...,T$.

\section{Multi-Period Portfolio Choice Problem for a Quadratic Utility Function: With Riskless Asset}

\noindent In this section we present a closed-form solution of the multi-period portfolio selection problem with $k$ risky assets and one riskless asset. Let
$\bx_t$ be the random return vector containing $k$ risky assets as defined in Section 2 and let $r_{f,t}$ be the return of the riskless asset at time $t$. Let $\bw_t=\left(w_{t,1},w_{t,2},\ldots,w_{t,k}\right)^{\prime}$ denote the vector of portfolio weights in period $t$ invested into the $k$ risky assets. Then the evolution of the investor's wealth is expressed as
\begin{equation}\label{wealth_Sec3}
W_t=W_{t-1}\left(1+r_{f,t}+\bw_{t-1}^{\prime}(\bx_t-r_{f,t}\bii)\right)=W_{t-1}\left(R_{f,t}+\bw^{\prime}_{t-1}\brx_t\right) \,,
\end{equation}
where $R_{f,t}=1+r_{f,t}$ and $\brx_{t}=\bx_{t}-r_{f,t}\bii$. The conditional mean vector of $\brx_t$ is $\brm_{t}=\bmu_t-r_{f,t}\bii$ and its conditional covariance matrix is given by $\bSigma_t$. In this section we consider an investor who invests into $k$ risky assets and one riskless asset with the investment strategy based on the quadratic utility function given in (\ref{QUF}). The optimization problem is given by
\begin{equation}\label{OP_Sec3}
V(0,W_0,\mathcal{F}_{0})=\max\limits_{\{\bw_s\}_{s=0}^{T-1}}E_t[U(W_T)]\,.
\end{equation}
Let
\begin{equation}\label{tF_t}
V(t,W_t,\mathcal{F}_{t})=\max\limits_{\{\bw_s\}_{s=t}^{T-1}}E_t[U(W_T)]\,.
\end{equation}

The problem (\ref{OP_Sec3}) is solved recursively by applying the following Bellman equation at time point $T-t$
\begin{eqnarray}\label{BE_Sec3}
&&V(T-t,W_{T-t},\mathcal{F}_{T-t})\nonumber\\
&=&\max\limits_{\bw_{T-t}}E_{T-t}
\Big{[}\max\limits_{\{\bw_s\}_{s=T-t+1}^{T-1}}E_{T-t+1}[U(W_T)]\Big{]} \\
&=&\max\limits_{\bw_{T-t}}E_{T-t}\Big{[}V(T-t+1,W_{T-t}\left(r_{f,T-t}+\bw^{*\;\prime}_{T-t+1}\brx_{T-t+1}\right),\mathcal{F}_{T-t+1})\Big{]}\nonumber
\end{eqnarray}
for $t=1,\ldots,T$. Note that for $t=0$ the equality (\ref{TC}) holds.

As in Section 2, we first derive the closed-form solution for the weights at period $T-1$. After that the general solution is derived (see Theorem 2). At time point $T-1$ the application of (\ref{TC}) leads to
\begin{equation}\label{VT}
V(T,W_T,\mathcal{F}_T)=U(W_T)=W_T-\frac{\alpha}{2}W_T^2\,.
\end{equation}

Let $\brA_{T}=\bSigma_{T}+\brm_{T}\brm^{\prime}_{T}$. Because
{\footnotesize
\begin{eqnarray*}
&&E_{T-1}\left(W_{T-1}(R_{f,T}+\bw^{\prime}_{T-1}\brx_{T})\right)^2= \Var_{T-1}\left(W_{T-1}(r_{f,T}+\bw^{\prime}_{T-1}\brx_{T})\right)\\
&+&\left(E_{T-1}\left(W_{T-1}(R_{f,T}+\bw^{\prime}_{T-1}\brx_{T})\right)\right)^2=W^2_{T-1}\Big{[}\bw^{\prime}_{T-1}\bA_{T}\bw_{T-1}+R^2_{f,T}+2R_{f,T}\bw^{\prime}_{T-1}\brm_{T}\Big{]}
\end{eqnarray*}
}
we get
{\footnotesize
\begin{eqnarray*}
&&V(T-1,W_{T-1},\mathcal{F}_{T-1})=\max\limits_{\{\bw_{T-1}\}}E_{T-1}\Big{[}E_T[W_T-\frac{\alpha}{2}W_T^2\Big{]}\\
&=&\max\limits_{\bw_{T-1}}E_{T-1}\Big{[}W_{T-1}\left(R_{f,T}+\bw^{\prime}_{T-1}\brx_{T}\right)
-\frac{\alpha}{2}\left(W_{T-1}(R_{f,T}+\bw^{\prime}_{T-1}\brx_{T}\right)^2\Big{]}\\
&=&\max\limits_{\bw_{T-1}}\Big{[}W_{T-1}\left(R_{f,T}+\bw^{\prime}_{T-1}\brm_{T}\right)
-\frac{\alpha}{2}E_{T-1}\left(W_{T-1}(R_{f,T}+\bw^{\prime}_{T-1}\brx_{T})\right)^2\Big{]}\\
&=&\max\limits_{\bw_{T-1}}\Big{[}W_{T-1}\left(R_{f,T}+\bw^{\prime}_{T-1}\brm_{T}\right)-\frac{\alpha}{2}W^2_{T-1}
\left(\bw^{\prime}_{T-1}\bA_{T}\bw_{T-1}+R^2_{f,T}+2R_{f,T}\bw^{\prime}_{T-1}\brm_{T}\right)
\Big{]}\,.
\end{eqnarray*}
}

The first order condition for the period $T-1$ is
\begin{equation}\label{FOC_Sec3}
W_{T-1}\brm_{T}-\alpha W^2_{T-1}\left(\brA_{T}\bw_{T-1}+R_{f,T}\brm_{T}\right)=0 \,.
\end{equation}
This leads to the following expression of the portfolio weights for the period $T-1$
\begin{equation}\label{weights_T-1_Sec3}
\bw^*_{T-1}=\left(\frac{1}{\alpha W_{T-1}}-R_{f,T}\right)\brA^{-1}_T\brm_T \,.
\end{equation}

In Theorem 2, the optimal multi-period portfolio weights are presented. The proof of the theorem is given in the appendix.

\begin{theorem}
Let $\bx_{\tau}=\left(X_{\tau,1},X_{\tau,2},\ldots,X_{\tau,k}\right)^{\prime}$, $\tau=0,\ldots,T$, be a random return vector of $k$ risky assets, let $r_{f,\tau}$ be the return of the riskless asset, and let $E(\bx_{\tau}|\mathcal{F}_{\tau-1})=\bmu_{\tau}$ and $Var(\bx_{\tau}|\mathcal{F}_{\tau-1})=\bSigma_{\tau}$ where $\bSigma_{\tau}$ is positive definite. Then for all $t=1,\ldots,T$ the optimal multi-period portfolio weights for period $T-t$ are given by
\begin{equation}\label{weights_T-t_Sec3}
\bw^*_{T-t}=\Big{[}\frac{1}{\alpha W_{T-t}}\left(\prod\limits_{i=T-t+2}^{T}R_{f,i}\right)^{-1}-R_{f,T-t+1}\Big{]}\brA^{-1}_{T-t+1}\brm_{T-t+1}^*\,
\end{equation}
with
\begin{equation}\label{brvA}
\brA_{T-t+1}= \left\{
  \begin{array}{l l}
    \bSigma_{T}+\brm_{T}\brm_{T}^\prime &\quad\text{for}~~~~t=1\\
    E_{T-t}[(1-\tilde{s}_{T-t+2})\brx_{T-t+1}\brx_{T-t+1}^\prime]& \quad \text{for}~~~ t=2,\ldots,T\\
  \end{array} \right.,
\end{equation}
\begin{equation}\label{brvmu}
\brm_{T-t+1}^*= \left\{
  \begin{array}{l l}
    \brm_{T} &\quad\text{for}~~~~t=1\\
    E_{T-t}\left((1-\tilde{s}_{T-t+2})\brx_{T-t+1}\right)& \quad \text{for}~~~ t=2,\ldots,T\\
  \end{array} \right.
\end{equation}
and
\begin{equation}\label{tildes}
\tilde{s}_{T-t+2}=\brm_{T-t+2}^{*\;\prime} \brA^{-1}_{T-t+2}\brm_{T-t+2}^* \quad \text{for}~~~ t=2,\ldots,T\,.
\end{equation}
\end{theorem}

As in the case without a riskless asset the expression of the optimal portfolio weights for each period $T-t$ looks like the solution of the single-period portfolio selection problem at time point $T-t$. In Corollary 2, a stronger result under the assumption of independence is proved, namely that the obtained weights are proportional to the weights obtained by solving \begin{equation}\label{SPP_Sec3}
\max \quad \bmu_t^\prime \bw-\frac{\alpha}{2}\bw^\prime \bSigma_t \bw \,.
\end{equation}

\begin{corollary}
Let $\bx_{\tau}=\left(X_{\tau,1},X_{\tau,2},\ldots,X_{\tau,k}\right)^{\prime}$, $\tau=0,\ldots,T$, be a sequence of the independently distributed vectors of $k$ risky assets, let $r_{f,\tau}$ be the return of the riskless asset, and let $E(\bx_{\tau})=\bmu_{\tau}$ and $Var(\bx_{\tau})=\bSigma_{\tau}$ where $\bSigma_{\tau}$ is positive definite. Then for all $t=1,\ldots,T$ the optimal multi-period portfolio weights for period $T-t$ are given by
\begin{equation}\label{weights_T-t_Sec3a}
\bw^*_{T-t}=\alpha_{T-t+1}^{-1}\bSigma^{-1}_{T-t+1}\brm_{T-t+1}\,,
\end{equation}
where
\begin{equation}\label{alph_T-t+1_Sec3}
\alpha_{T-t+1}= \frac{\Big{[}\frac{1}{\alpha W_{T-t}}\left(\prod\limits_{i=T-t+2}^{T}R_{f,i}\right)^{-1}-R_{f,T-t+1}\Big{]}}{1+\brm_{T-t+1}^\prime \bSigma^{-1}_{T-t+1}\brm_{T-t+1}}\,.
\end{equation}
\end{corollary}

Corollary 2 shows that the solution of (\ref{OP_Sec3}) at each time point $T-t$ is similar to the solution of the common single-period optimization problem given in (\ref{SPP_Sec3}). The only difference is that the coefficient $\alpha_{T-t}$ is time varying. It depends on the future returns of the riskless asset. Leippold et al. (2004) suggested a similar solution to one presented in Corollary 2 by using the geometric approach in the case of independent asset returns.

It appears that the expressions presented in Theorem 2 are quite complicate and cannot be evaluated for an arbitrary model of the asset returns. For this reason we suggest a reasonable approximation for the weights and evaluate its accuracy.

First, the moments presented in (\ref{brvA}) and (\ref{brvmu}) are approximated. Let $\xi$ be a random variable with support $[0,1]$. The application of the Sherman-Morrison formula to $1-\tilde{s}_{T-t+2}$ leads to
\begin{equation}\label{sec4_eq1}
1-\xi\tilde{s}_{T-t+2}=\frac{1+(1-\xi)\brm_{T-t+1}^\prime\bSigma_{T-t+1}^{-1}\brm_{T-t+1}}{1+\brm_{T-t+1}^\prime\bSigma_{T-t+1}^{-1}\brm_{T-t+1}}=\frac{1+(1-\xi)\breve{s}_{T-t+2}}{1+\breve{s}_{T-t+2}} \,,
\end{equation}
where $\breve{s}_{T-t+2}=\brm_{T-t+2}^\prime\bSigma_{T-t+2}^{-1}\brm_{T-t+2}$. Let $\eta_{T-t+2}=\dfrac{1+(1-\xi)\breve{s}_{T-t+2}}{1+\breve{s}_{T-t+2}}$ whose support is $[0,1]$ as well.
Then, it holds for the $i$-th component of $\brm^*_{T-t+1}$ that
\begin{eqnarray*}
&&E_{T-t}\left((1-\xi \tilde{s}_{T-t+2})\brx_{T-t+1,i}\right)=E_{T-t}\left(\frac{1+(1-\xi)\breve{s}_{T-t+2}}{1+\breve{s}_{T-t+2}}\brx_{T-t+1,i}\right)\\
&=&E_{T-t}\left(\eta_{T-t+2}\brx_{T-t+1,i}\right)\\
&=&E_{T-t}\left(\eta_{T-t+2}\left(\brx_{T-t+1,i}-\brm_{T-t+1,i}+\brm_{T-t+1,i}\right)\right)\\
&=&E_{T-t}\left(\eta_{T-t+2} \brm_{T-t+1,i}\right)+E_{T-t}\left(\eta_{T-t+2}(\brx_{T-t+1,i}-\brm_{T-t+1,i})\right)\\
&=&E_{T-t}\left(\eta_{T-t+2}\right)\brm_{T-t+1,i}+E_{T-t}\left(\eta_{T-t+2}(\brx_{T-t+1,i}-\brm_{T-t+1,i})\right)\,.
\end{eqnarray*}

Because $\eta_{T-t+2}\le 1$ it holds
{
\begin{eqnarray*}
\left|E_{T-t}\left(\eta_{T-t+2}(\brx_{T-t+1,i}-\brm_{T-t+1,i})\right)\right|&\le& E_{T-t}\left(\left|\brx_{T-t+1,i}-\brm_{T-t+1,i}\right|\right)\le\\
&\le& \sqrt{MSE_{T-t+1,i}}\,,
\end{eqnarray*}
}
where $MSE_{T-t+1,i}=E_{T-t}\left((\brx_{T-t+1,i}-\brm_{T-t+1,i})^2\right)$ is the conditional mean square prediction error calculated for $\brm_{T-t+1,i}$. If $MSE_{T-t+1,i}$ is small, what should be expected if a good forecast of the process future values is performed, then the quantity $\brm^*_{T-t+1,i}$ is well approximated by
\begin{equation}\label{approx_brm*}
\brm^*_{T-t+1,i} \approx E_{T-t}\left(\eta_{T-t+2}\right)\brm_{T-t+1,i} \,.
\end{equation}

Similar results are obtained for (\ref{brvA}). Here, it holds that
{\footnotesize
\begin{eqnarray*}
&&E_{T-t}\left((1-\tilde{s}_{T-t+2})\brx_{T-t+1,i}\brx_{T-t+1,j}\right)=E_{T-t}\left(\eta_{T-t+2}\brx_{T-t+1,i}\brx_{T-t+1,j}\right)\\
&=&E_{T-t}\left(\eta_{T-t+2}(\brx_{T-t+1,i}-\brm_{T-t+1,i}+\brm_{T-t+1,i})(\brx_{T-t+1,j}-\brm_{T-t+1,j}+\brm_{T-t+1,j})\right)\\
&=&E_{T-t}\left(\eta_{T-t+2}\brm_{T-t+1,i}\brm_{T-t+1,j}\right)\\
&+&E_{T-t}\left(\eta_{T-t+2}(\brx_{T-t+1,i}-\brm_{T-t+1,i})(\brx_{T-t+1,j}-\brm_{T-t+1,j})\right)\\
&+&\brm_{T-t+1,i}E_{T-t}\left(\eta_{T-t+2}(\brx_{T-t+1,j}-\brm_{T-t+1,j})\right)\\
&+&\brm_{T-t+1,j}E_{T-t}\left(\eta_{T-t+2}(\brx_{T-t+1,i}-\brm_{T-t+1,i})\right)\\
&=&(b_{T-t+1,ij}-\sigma_{T-t+1,ij})E_{T-t}\left(\eta_{T-t+2}\right)\\
&+&E_{T-t}\left(\eta_{T-t+2}(\brx_{T-t+1,i}-\brm_{T-t+1,i})(\brx_{T-t+1,j}-\brm_{T-t+1,j})\right)\\
&+&\brm_{T-t+1,i}E_{T-t}\left(\eta_{T-t+2}(\brx_{T-t+1,j}-\brm_{T-t+1,j})\right)\\
&+&\brm_{T-t+1,j}E_{T-t}\left(\eta_{T-t+2}(\brx_{T-t+1,i}-\brm_{T-t+1,i})\right)\,,
\end{eqnarray*}
where $\mathbf{B}_{T-t+1}=E_{T-t}(\brx_{T-t+1}\brx_{T-t+1}^\prime)=(b_{T-t+1,ij})_{i,j=1,...k}$ and $\mathbf{\Sigma}_{T-t+1}=(\sigma_{T-t+1,ij})_{i,j=1,...k}$.}

Because $0 \le \eta_{T-t+2}\le 1$ we get that
{\footnotesize \begin{eqnarray*}
&&\left|E_{T-t}\left(\eta_{T-t+2}(\brx_{T-t+1,i}-\brm_{T-t+1,i})(\brx_{T-t+1,j}-\brm_{T-t+1,j})\right)\right.\\
&-&\left.\sigma_{T-t+1,ij}E_{T-t}\left(\eta_{T-t+2}\right)+\brm_{T-t+1,i}E_{T-t}\left(\eta_{T-t+2}(\brx_{T-t+1,j}-\brm_{T-t+1,j})\right)\right.\\
&+&\left.\brm_{T-t+1,j}E_{T-t}\left(\eta_{T-t+2}(\brx_{T-t+1,i}-\brm_{T-t+1,i})\right)\right|\\
&\le& E_{T-t}\left(|\brx_{T-t+1,i}-\brm_{T-t+1,i}| |\brx_{T-t+1,j}-\brm_{T-t+1,j}|\right)+\left|\sigma_{T-t+1,ij}\right|\\
&+&|\brm_{T-t+1,i}|E_{T-t}\left(|\brx_{T-t+1,j}-\brm_{T-t+1,j}|\right)+|\brm_{T-t+1,j}|E_{T-t}\left(|\brx_{T-t+1,i}-\brm_{T-t+1,i}|\right)\\
&\le& \sqrt{MSE_{T-t+1,i}}\sqrt{MSE_{T-t+1,j}}+\left|\sigma_{T-t+1,ij}\right|\\
&+&|\brm_{T-t+1,i}|\sqrt{MSE_{T-t+1,j}}+|\brm_{T-t+1,j}|\sqrt{MSE_{T-t+1,i}}\,.
\end{eqnarray*}
}
Furthermore, we point out that
{\footnotesize \begin{equation*}
|\sigma_{T-t+1,ij}|\le \sqrt{\sigma_{T-t+1,ii}}\sqrt{\sigma_{T-t+1,jj}}\le \sqrt{MSE_{T-t+1,i}} \sqrt{MSE_{T-t+1,j}} .
\end{equation*}

}
Hence, if $MSE_{T-t+1,i}$ are relatively small for all $i=1,...,k$ we get the following approximation
{
\begin{eqnarray}
\label{approx_brA*}
\breve{\bA}_{T-t+1,i} &\approx& E_{T-t}\left(\eta_{T-t+2}\right) E_{T-t}(\brx_{T-t+1}\brx_{T-t+1}^\prime) \nonumber\\
&=&E_{T-t}\left(\eta_{T-t+2}\right)\left(\bSigma_{T-t+1}+\brm_{T-t+1}\brm_{T-t+1}^\prime\right)\,.
\end{eqnarray}
}
Finally, we note that $\eta_T=1$ and $\eta_{T-t+1}=\dfrac{1+(1-E_{T-t}(\eta_{T-t+2}))\breve{s}_{T-t+1}}{1+\breve{s}_{T-t+1}}$ for $t \ge 2$.

Putting (\ref{approx_brm*}) and (\ref{approx_brA*}) together we obtain the following approximation for the weights
{\small
\begin{equation}\label{weights_T-t_Sec3_app}
\bw^*_{T-t}=\Big{[}\frac{1}{\alpha W_{T-t}}\left(\prod\limits_{i=T-t+2}^{T}R_{f,i}\right)^{-1}-R_{f,T-t+1}\Big{]}\left(\bSigma_{T-t+1}+\brm_{T-t+1}\brm_{T-t+1}^\prime\right)^{-1}
\brm_{T-t+1}\,.
\end{equation}
}
The approximation (\ref{weights_T-t_Sec3_app}) can be used for arbitrary models of the asset returns
for which $MSE_{T-t+1,i}, i=1,...,k$ is relatively small. This is closely related to the problem that the considered stochastic model for the asset returns provides a good fit to real data. As a result, we conclude that if an appropriate stochastic model is chosen the suggested approximation for the multi-period optimal portfolio weights works well. Moreover, this approximation becomes an exact one if the asset returns are independently distributed as it is shown in Corollary 2.

Very interesting results can be obtained for the tangency portfolio as well (see, e.g. Ingersoll (1987, p. 89), Britten-Jones (1999)). The weights of the tangency portfolio are derived at each time point $T-t$ from (\ref{weights_T-t_Sec3a}) by assuming that the whole wealth is invested only into the risky assets, i.e. under the assumption $\bw_{T-t}^\prime \bii=1$. The results for the multi-period portfolio allocation problem with the tangency portfolio under the assumption of independence are given in Theorem 3.

\begin{theorem}
Let $\bx_{\tau}=\left(X_{\tau,1},X_{\tau,2},\ldots,X_{\tau,k}\right)^{\prime}$, $\tau=0,\ldots,T$, be a sequence of the independently distributed vectors of $k$ risky assets, let $r_{f,\tau}$ be the return of the riskless asset, and let $E(\bx_{\tau})=\bmu_{\tau}$ and $Var(\bx_{\tau})=\bSigma_{\tau}$ where $\bSigma_{\tau}$ is positive definite. Then for all $t=1,\ldots,T$ the tangency portfolio weights for period $T-t$ are given by
\begin{equation}\label{weights_T-t_TP}
\bw^*_{TP,T-t}=\frac{\brA^{-1}_{T-t+1}\brm_{T-t+1}}{\bii^\prime \brA^{-1}_{T-t+1}\brm_{T-t+1}}=
\frac{\bSigma^{-1}_{T-t+1}\brm_{T-t+1}}{\bii^\prime \bSigma^{-1}_{T-t+1}\brm_{T-t+1}}\,.
\end{equation}
\end{theorem}

Theorem 3 shows that for the tangency portfolio the multi-period portfolio selection problem is equivalent to the single-period allocation problem solved at each time point $T-t$. Because the tangency portfolio is, usually, considered as a market portfolio in the single-period allocation problem (see, e.g., Britten-Jones (1999)), it can also be treated as a benchmark portfolio in the multi-period case, provided that the asset returns are independent.

\section{Empirical Illustration}

\noindent In this part of the paper we apply the results of Section 3 to real data. In the first example we deal with the multi-period portfolio selection under return predictability. In this subsection a model presented by Brandt and Santa-Clara (2006) is applied and approximative expressions to those presented in Theorem 1 are analyzed. In the second empirical illustration we consider an investor who attends to invest into the international portfolio consisting of five developed stock markets, namely Belgium, Germany, Japan, the UK, and the USA. The approximative solutions of multi-period portfolio selection problem are obtained in this case as well.

The investment strategy based on equation (\ref{weights_T-t_Sec3_app}) will be called as the investment strategy based on the predictive loss approximation with a riskless asset and it is briefly denoted as the LAMPS strategy. It is noted that the LAMPS strategy coincides with the expressions given in Corollary 2 that are derived for the multi-period portfolio selection problem with riskless asset under the assumption that the asset returns are independent. Similar results are obtained in the case of the multi-period tangency portfolio (MTP). The application of the predictive loss approximation leads to the formulas presented in Theorem 2.

As a first benchmark strategy of the multi-period portfolio selection we consider an investor who chooses the global minimum variance (GMV) portfolio which in the present situation is calculated by the time-invariant weights expressed as
\begin{equation}\label{GMV_w}
\bw_{GMV,t}=\frac{\bSigma^{-1}\bii}{\bii^\prime \bSigma^{-1}\bii} \quad \text{for} \quad t=1,...,T\,.
\end{equation}
Because the same proportions of the wealth are invested into each stock at each time point we consider this portfolio as a myopic strategy.

We also compare the results obtained by using the predictive loss approximation to the expression of the weights presented in Theorem 2 with two other benchmark strategies in the case with a riskless asset. The first strategy is known as a partial myopic strategy, i.e. the whole wealth is invested into the riskless asset. This strategy was suggested by Mossin (1968) who derived conditions under the utility function under which this strategy is optimal in the case of the portfolio consisting of one stock and the riskless asset. A further considered benchmark method is one suggested by Brandt and Santa Clara (2006) that is based on the assumption that
\begin{equation}\label{weights_BS}
\bw_t=\bol{\theta} \bz_t\,,
\end{equation}
where $\bz_t$ is the vector of predictable variables taken at time point $t$. We refer to this portfolio strategy as the BSC strategy. In the empirical illustration, Brandt and Santa Clara (2006) argued that the solution of the multi-period portfolio selection problem based of this approach is very close to the one obtained from the simulation approach applied to the dynamic portfolio choice problem as suggested by Brandt et al. (2005) and van Binsbergen and Brandt (2007).

\subsection{Multi-Period Portfolio Selection under Return Predictability}

\noindent In this section we deal with the multi-period portfolio selection problem assuming that the asset returns are predictable. This is one of the most commonly used approach applied for modeling the time series properties of the asset returns (see, e.g. Campbell and Viceira (2002), Brandt and Santa Clara (2006)). Here, we consider a special case of a model  suggested by Brandt and Santa Clara (2006). They applied it to monthly returns from January 1945 to December 2000 of one stock ($r_t^s$) and one bond ($r_t^b$). The term spread is used as a predictable variable $z_t$. Brandt and Santa Clara (2006, p.2200) obtain the model
\begin{equation}\label{VAR_BS}
\left[\begin{array}{c}
\ln(1+r_{t+1}^s) \\
\ln(1+r_{t+1}^b)\\
z_{t+1}\\
\end{array}\right]=\left[
\begin{array}{r}
0.0059 \\
0.0007\\
-0.0028\\
\end{array}\right]+\left[
\begin{array}{r}
0.0060 \\
0.0035\\
0.9597\\
\end{array}\right]\times z_t+\left[
\begin{array}{r}
\varepsilon^s_{t+1} \\
\varepsilon^b_{t+1} \\
\varepsilon^z_{t+1} \\
\end{array}\right]
\end{equation}
with
\begin{equation}\label{Eps_distr}
\left[
\begin{array}{r}
\varepsilon^s_{t+1} \\
\varepsilon^b_{t+1} \\
\varepsilon^z_{t+1} \\
\end{array}\right]\sim MVN
\left[
\begin{array}{c}
\\
0\\
\\
\end{array},\left[
\begin{array}{rrr}
0.0018&0.0002&-0.0005\\
0.0002&0.0006&0.0007\\
-0.0005&0.0007&0.0802\\
\end{array}\right]
\right].
\end{equation}

Note that the model (\ref{VAR_BS}) can be presented as a vector autoregressive process of order $1$. Let $\mathbf{Y}_t=(\ln(1+r_{t+1}^s),\ln(1+r_{t+1}^b), z_t)^\prime$ then (\ref{VAR_BS}) becomes
\begin{equation}\label{VAR}
\by_t=\tilde{\bn}+\tbF\by_{t-1}+\tilde{\eps}_t\,
\end{equation}
with $\tilde{\eps}_t\sim \mathcal{N}(\bzero, \tilde{\bSigma})$.
Thus, for $\mathbf{X}_t$ we get an autoregressive representation with
\begin{equation}\label{VAR11}
\bx_t=\bL\tilde{\bn}+\bL\tbF\by_{t-1}+\bL\tilde{\eps}_t=\bn+\bF\by_{t-1}+\eps_{t}~~~\text{with}~~\bL=[\bI~ \bzero]\,,
\end{equation}
where $\bI$ is a $2\times 2$ indentity matrix and $\bzero$ is a $2\times 1$ vector of zeros.
Consequently, $\bx_t|\mathcal{F}_{t-1}\sim \mathcal{N}(\bmu_t, \bSigma)$. We are interested in the conditional mean vector and in the conditional covariance matrix of $\mathbf{X}_t$ given $\mathcal{F}_{t-1}$. Note that
\begin{equation}\label{cova}
\bmu_t=E(\bx_t|\mathcal{F}_{t-1})=\bn+\bF\by_{t-1},~~ \bSigma=\Var(\bx_t|\mathcal{F}_{t-1})=\bL\tilde{\bSigma}\bL^\prime\,,
\end{equation}
 Inserting (\ref{cova}) in (\ref{weights_T-t_Sec3_app}) leads to the weights of the LAMPS strategy. In similar way the weights of the other multi-period portfolio strategies are calculated.

We compare the performance of the above derived strategies with each other via an extensive simulation study based on $10^5$ independent repetitions. Multi-period portfolio strategies are constructed for $T \in \{6,12,18,24\}$ and $\alpha \in \{0.833,0.909,\\ 0.937,0.952\}$. The values of $\alpha$ correspond to $\gamma \in \{5,10,15,20\}$ which are also used in Brandt and Santa Clara (2006).

\begin{center}
\begin{table}[ptbh]
\begin{center}
\begin{tiny}
\begin{tabular}{|c|c|c|c|c|c|}
\hline
%\multicolumn{6}{|c|}{Monthly Data with predictable variable (Brandt) - \textbf{Expected Utility}}\\
%\hline
\backslashbox{T}{$\gamma$} &5&10&15&20& Method\\ \hline

\multirow{4}{*}{6}
& \textbf{0.5904} (0.0061) & \textbf{0.5474} (0.0016) & \textbf{0.5324} (0.0008) & \textbf{0.5245} (0.0004) & LAMPS\\
& 0.5847 (0.0065) & 0.5461 (0.0031) & 0.5317 (0.0017) & 0.5239 (0.0012) & GMV\\
& 0.5837 & 0.5456 & 0.5316 & 0.5241 & Part.Myopic\\
& 0.5834 (0.0157) & 0.5459 (0.0039) & 0.5317 (0.0017) & 0.5242 (0.0010)& BSC\\

 \hline

\multirow{4}{*}{12}
& \textbf{0.5949} (0.0046)  & \textbf{0.5487} (0.0012)   & \textbf{0.5330} (0.0006)  & \textbf{0.5249} (0.0003) & LAMPS\\
& 0.5858 (0.0092)  & 0.5460 (0.0035)   & 0.5310 (0.0024)  & 0.5229 (0.0021) & GMV\\
& 0.5839 & 0.5457 & 0.5316 & 0.5241 &Part.Myopic\\
& 0.5822 (0.0172)  & 0.5455 (0.0043)   & 0.5316 (0.0019)  & 0.5241 (0.0011) & BSC\\

 \hline

\multirow{4}{*}{18}
& \textbf{0.5975 }(0.0027) &  \textbf{0.5493} (0.0007) & \textbf{0.5333} (0.0003) & \textbf{0.5250} (0.0002) &LAMPS\\
& 0.5867 (0.0106) &  0.5452 (0.0044) & 0.5298 (0.0035) & 0.5218 (0.0032) & GMV\\
& 0.5841 & 0.5458 & 0.5317 & 0.5242 & Part.Myopic\\
& 0.5801 (0.0195) &  0.5450 (0.0049) & 0.5313 (0.0022) & 0.5239 (0.0012) &BSC\\

 \hline

\multirow{4}{*}{24}
& \textbf{0.5989} (0.0014) & \textbf{0.5497} (0.0004) & \textbf{0.5335} (0.0002) & \textbf{0.5251} ($8.5\cdot10^{-5}$) &LAMPS\\
& 0.5869 (0.0113) & 0.5441 (0.0054) & 0.5285 (0.0047) & 0.5204 (0.0044) & GMV\\
& 0.5843 & 0.5459 & 0.5318 & 0.5242 & Part.Myopic\\
& 0.5757 (0.0240) & 0.5441 (0.0058) & 0.5309 (0.0027) & 0.5237 (0.0015) &BSC\\

\hline
\end{tabular}
\end{tiny}
\end{center}
\caption{Medians and median absolute deviations (MAD, in parentheses) of the expected quadratic utility for the LAMPS, the multi-period GMV, the partial myopic, and the BSC strategies
in the case of data from the example of Brandt and Santa Clara (2006).
}
\label{tab1}%
\end{table}
\end{center}

In Table 1 we present the median of the expected utility functions for the considered values of $\gamma$ and the investment horizon $T$. In each block the strategies order is the
LAMPS, multi-period GMV, partial myopic, and BSC. In the parentheses the median absolute deviation (MAD) is given that is calculated as the median of the absolute deviation of
the sample values from the median. The monthly data from January, 1945 to December, 2000 from the example of Brandt and Santa Clara (2006) are used that consist of one stock and one bond return at each time point the portfolio, while the term structure is used as a predictable variable for modeling time series properties of the return process.

\begin{landscape}
\renewcommand{\baselinestretch}{1.0}
\begin{figure}
\caption{\footnotesize The empirical distribution function of the expected quadratic utility for the LAMPS, the multi-period GMV, the partial myopic, and the BSC strategies in the case of data from the example of Brandt and Santa Clara (2006). We consider $\gamma \in \{5,10,15,20\}$ and $T=6$.}
\vspace{-0.3cm}
\begin{center}
\includegraphics[angle=270, width=17.8cm]{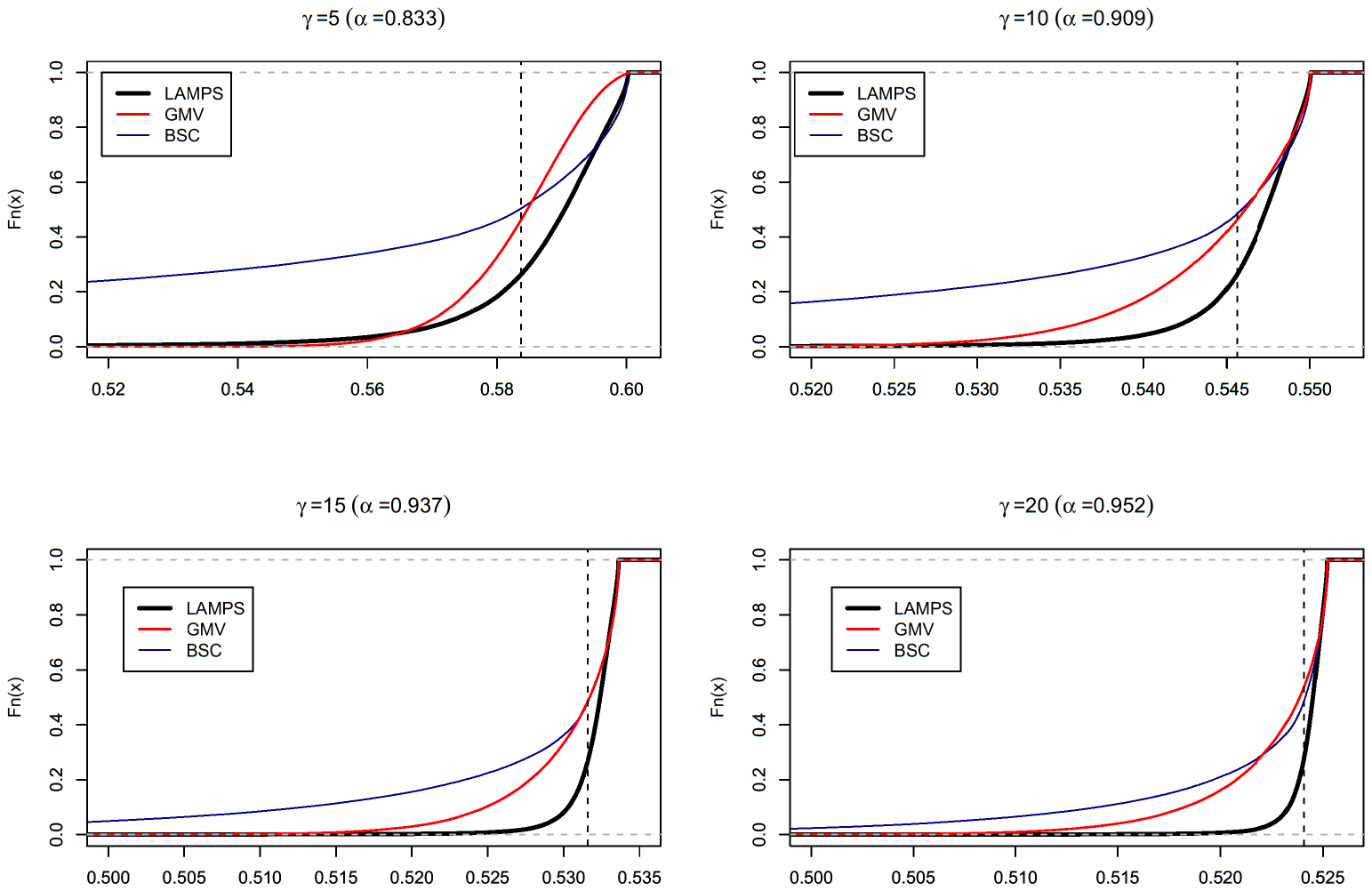}
\end{center}
\label{fig:ECDF_Brandt1}
\end{figure}
\end{landscape}

\newpage
\begin{landscape}
\renewcommand{\baselinestretch}{1.0}
\begin{figure}
\caption{\footnotesize The empirical distribution function of the expected quadratic utility for the LAMPS, the multi-period GMV, the partial myopic, and the BSC strategies in the case of data from the example of Brandt and Santa Clara (2006). We consider $\gamma \in \{5,10,15,20\}$ and $T=12$.}
\vspace{-0.3cm}
\begin{center}
\includegraphics[angle=270, width=17.8cm]{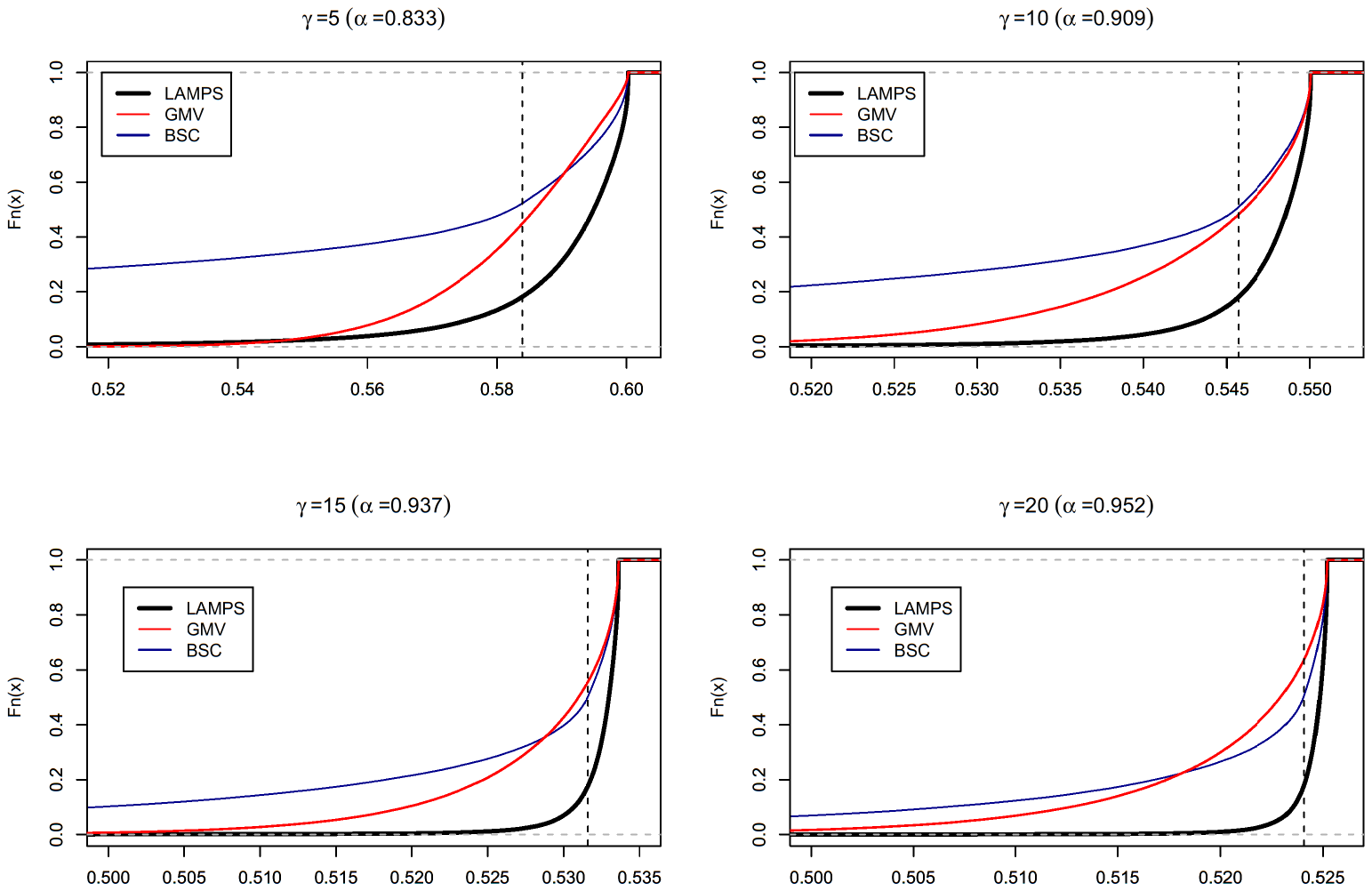}
\end{center}
\label{fig:ECDF_Brandt2}
\end{figure}
\end{landscape}

\newpage
\begin{landscape}
\renewcommand{\baselinestretch}{1.0}
\begin{figure}
\caption{\footnotesize The empirical distribution function of the expected quadratic utility for the LAMPS, the multi-period GMV, the partial myopic, and the BSC strategies in the case of data from the example of Brandt and Santa Clara (2006). We consider $\gamma \in \{5,10,15,20\}$ and $T=18$.}
\vspace{-0.3cm}
\begin{center}
\includegraphics[angle=270, width=17.8cm]{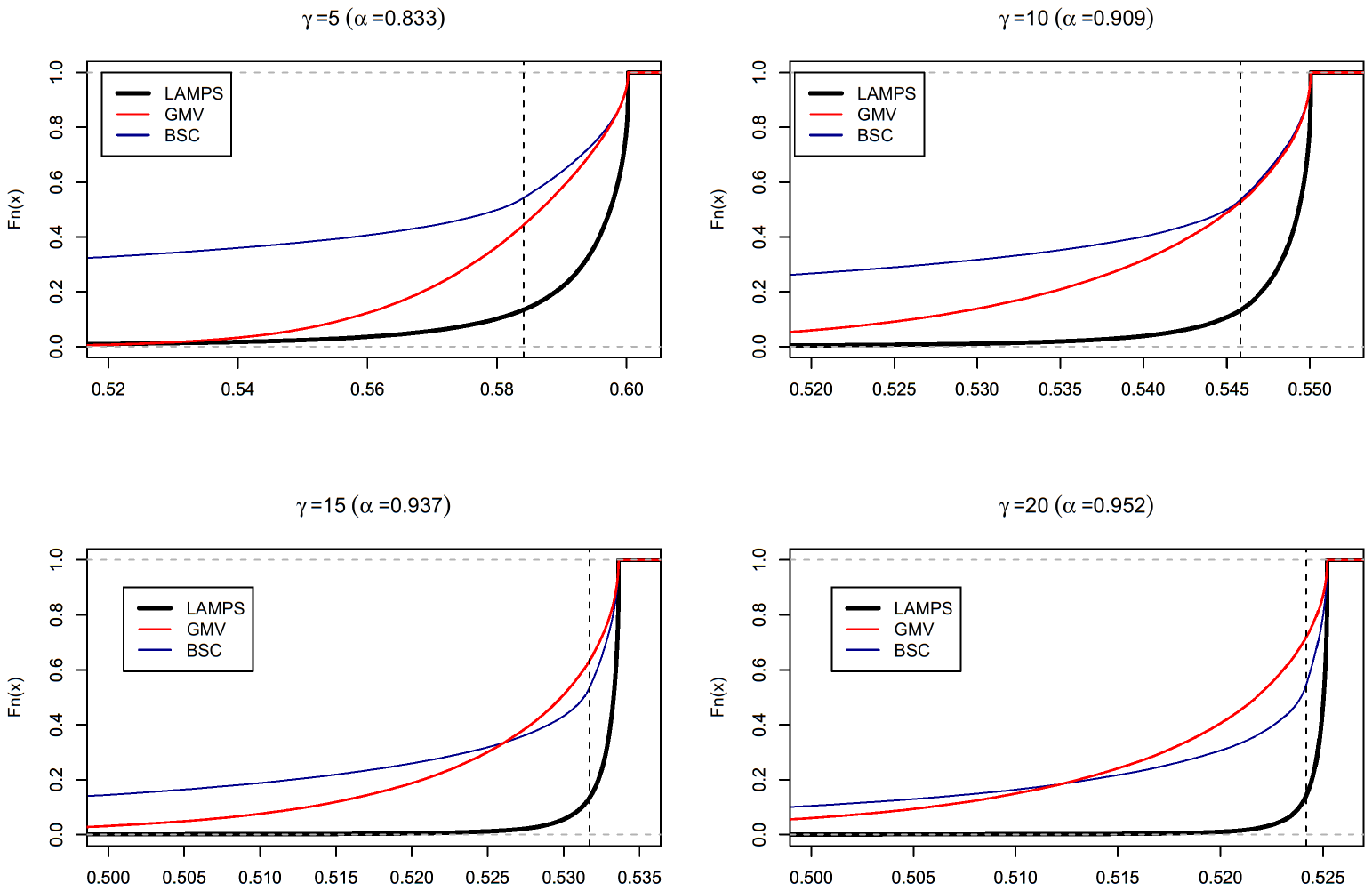}
\end{center}
\label{fig:ECDF_Brandt3}
\end{figure}
\end{landscape}

\newpage
\begin{landscape}
\renewcommand{\baselinestretch}{1.0}
\begin{figure}
\caption{\footnotesize The empirical distribution function of the expected quadratic utility for the LAMPS, the multi-period GMV, the partial myopic, and the BSC strategies in the case of data from the example of Brandt and Santa Clara (2006). We consider $\gamma \in \{5,10,15,20\}$ and $T=24$.}
\vspace{-0.3cm}
\begin{center}
\includegraphics[angle=270, width=17.8cm]{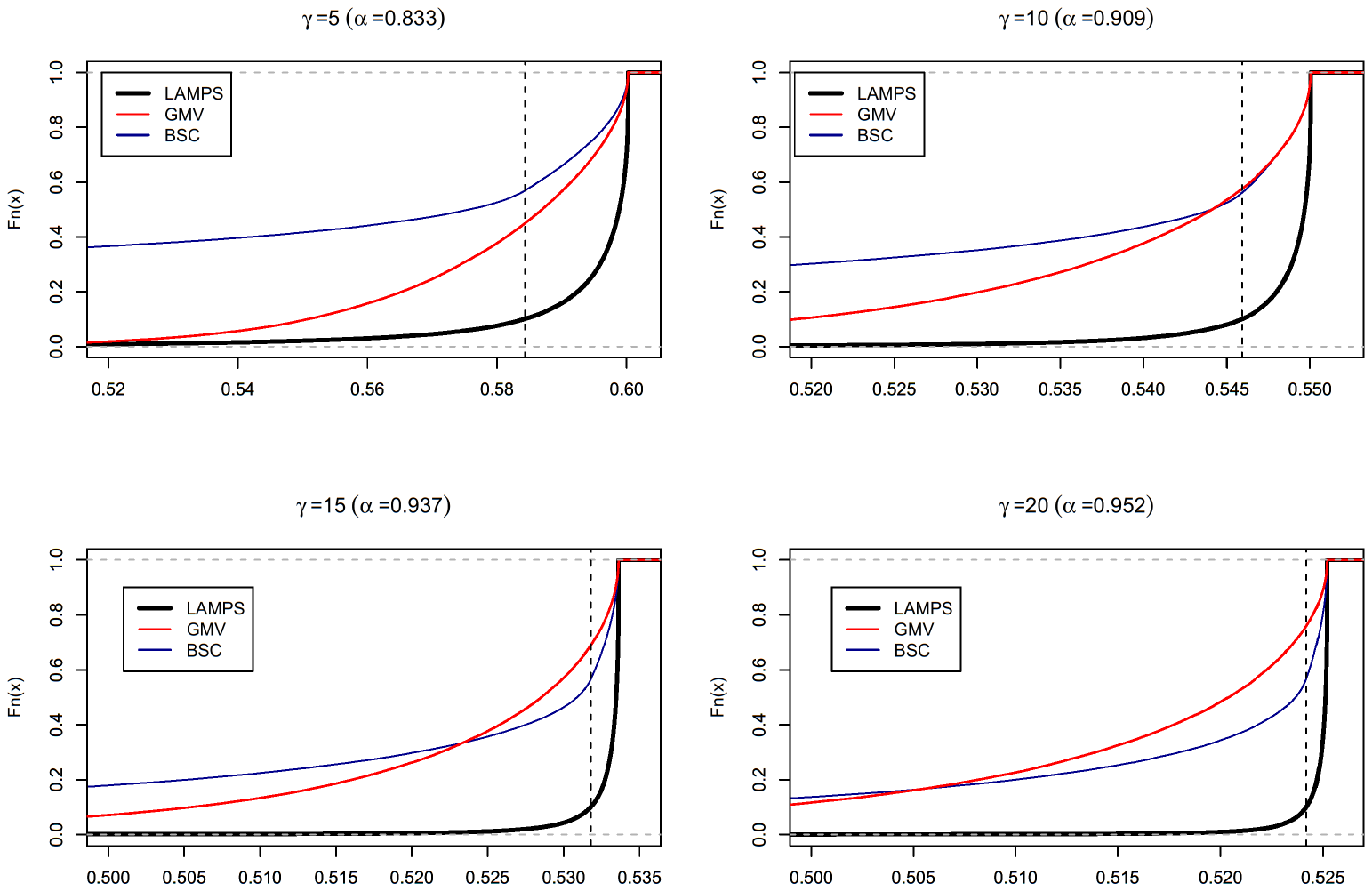}
\end{center}
\label{fig:ECDF_Brandt4}
\end{figure}
\end{landscape}

\newpage
\begin{landscape}
\renewcommand{\baselinestretch}{1.0}
\begin{figure}
\caption{\footnotesize The median, the $5\%$ and $95\%$ quantiles of the expected quadratic utility for the LAMPS strategy as a function of $T=\{4,8,12,16,20,24,28,32,36,40\}$ in the case of data from the example of Brandt and Santa Clara (2006). We consider $\gamma \in \{5,10,15,20\}$.}
\vspace{-0.3cm}
\begin{center}
\includegraphics[angle=0, width=17.8cm]{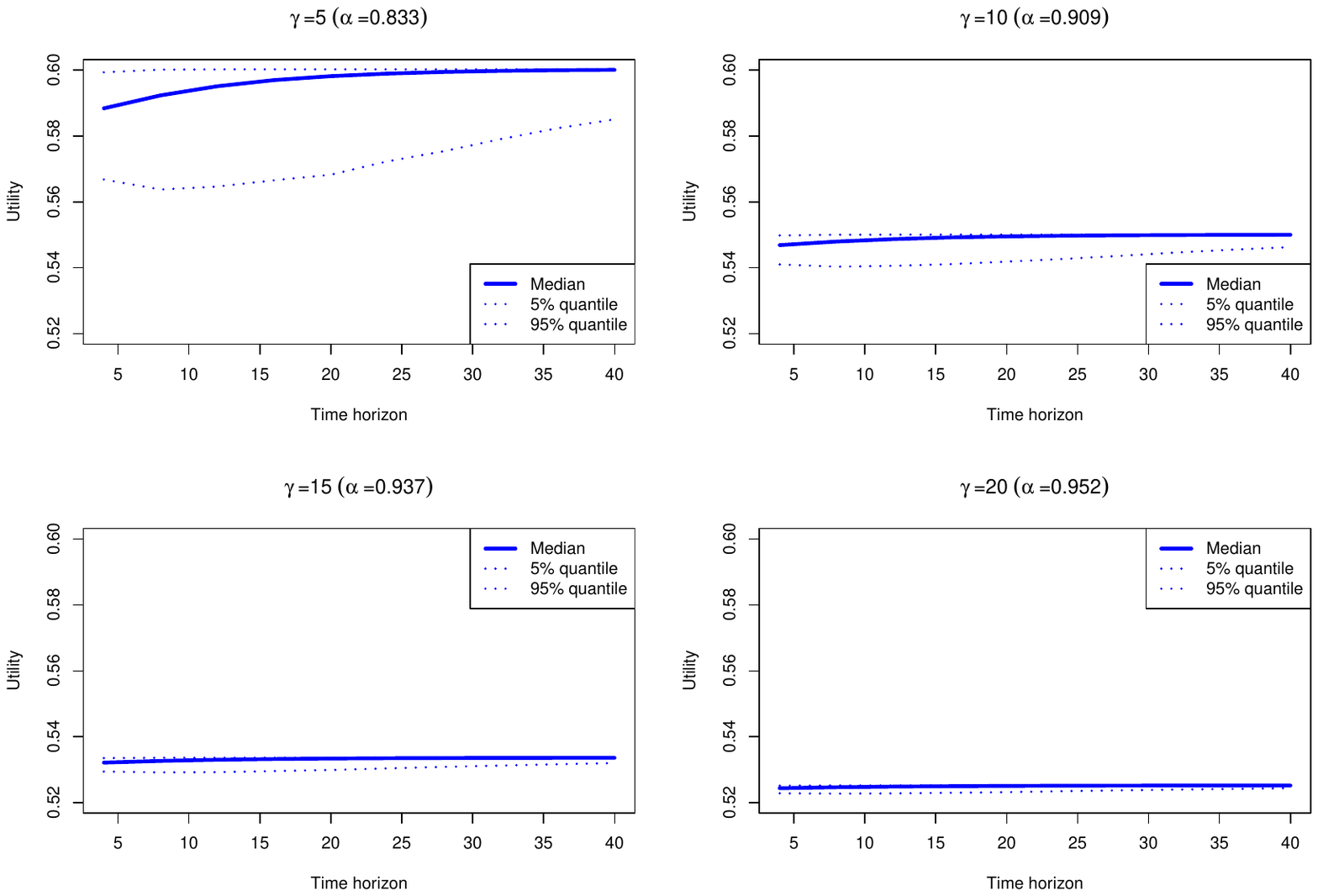}
\end{center}
\label{fig:ECDF_Brandt4b}
\end{figure}
\end{landscape}

We observe a very good performance of the LAMPS strategy which is the best strategy for all $T$ and $\gamma$.
On the other hand it is not possible to provide a clear ranking between the GMV, partial myopic, and BSC strategies.
While the GMV strategy is on the second place for smaller values of $\gamma$ and each $T$, the opposite results are observed for larger values of $\gamma$
where the partial myopic and the BSC strategies are ranked on the second and third places.
%On the last place the IAMPS is present for all $\gamma$ and $T$.
Moreover, we obtained very small values of the MAD for the LAMPS strategy, while the largest values are attained for the GMV approach.

In Figures 1-4 we study the problem in more details. Here, we present the empirical distribution functions of the expected quadratic utility functions obtained via the simulation study. If we compare several strategies based on the performance of their empirical distribution functions we should choose the strategy whose distribution function lies below the others. It follows from the fact that in this case for a fixed value of $W$ the probability to obtain stochastically larger value is the largest one, i.e. with the higher probability larger values of the expected utility are attained. Because the partial myopic strategy is based on investing into a riskless asset it is a deterministic strategy and it's distribution function is a step function. The position of the given index is used for comparing the dynamic portfolio strategies. This criterion is based on the maximum probability of exceeding the utility of the partial myopic portfolio.

In Figure 1 we observe a very good performance of the LAMPS and BSC strategies, especially for smaller values of $\gamma$ and $T$. If $\gamma$ increases then the probability of overperforming the partial myopic strategy decreases. It is always small for the GMV approach. For this method it is less than $40\%$ for larger values of $\gamma$ and $T$, while it is very large for the LAMPS strategy (about $80\%$).

In Figure 5 we analyse how quickly the resulting quadratic utility of the LAMPS strategy converges to its maximum value as the investment horizon $T$ increases. Here, we plot the median as well as $5\%$ and $95\%$ quantiles of the calculated expected quadratic utility for $T=\{4,8,12,16,20,24,28,32,36,40\}$ and $\gamma \in \{5,10,15,20\}$. Note that the median quickly converges to the corresponding $95\%$ quantile and that already for  moderate values of $T \ge 25$ a good approximation is obtained. The $5\%$ quantile also becomes considerably large as $T$ increases. Consequently, the results presented in the figure lead to the conclusion that the expected quadratic utility converges to its maximum value as the investment horizon becomes larger, although the convergence rate is not large, especially for smaller values of $\gamma$.

\subsection{Multi-Period International Optimal Portfolio}

\noindent We consider an investor who invests into an international portfolio consisting of the capital markets indices of five developed stock market, namely Belgium, Germany, Japan, the UK, and the USA. The weekly data of the MSCI (Morgan Stanley Capital International) indices for the equity markets returns are considered from the 4th of January, 2002 to the 4th of December, 2009 and the return series for each index are calculated. To the return series we fit the VAR(1)-CCC-GARCH(1,1) process defined by
\begin{equation}\label{VAR}
\bx_t=\boldsymbol{\nu}+\mathbf{\Phi}\by_{t-1}+\bSigma^{1/2}_t\eps_t \quad \text{with} \quad \eps_t \sim ii\mathcal{N}(\mathbf{0},\bI)\,
\end{equation}
and $ \bSigma_t=\text{diag}\{\mathbf{h}_t\}^{1/2}~\bR~\text{diag}\{\mathbf{h}_t\}^{1/2}$ with
\begin{equation}
\mathbf{h}_t=\mathbf{a}_0+\mathbf{A}\bol{\xi}_{t-1}+\mathbf{B}\mathbf{h}_{t-1}\,,
\end{equation}
where $\bol{\xi}_t=(\eps_{t,1}^2,...,\eps_{t,k}^2)^\prime$; $\mathbf{A}$ and $\mathbf{B}$ are diagonal matrices.

We obtain
\begin{equation}\label{VAR_int}
\bol{\nu}=
\left[
\begin{array}{r}
\operatorname{4.83e-04}\\
\operatorname{1.20e-03}\\
\operatorname{6.74e-04}\\
\operatorname{5.54e-04}\\
\operatorname{2.79e-05}\\
\end{array}\right], \mathbf{\Phi}=\left[
\begin{array}{rrrrr}
0.2011& \operatorname{-0.1592}&  0.01892& \operatorname{-0.196}& {\bf 0.455}\\
0.3139& \operatorname{-0.1231}& \operatorname{-0.00191}& \operatorname{-0.511}&\textbf{ 0.434}\\
0.0487&  0.0888& \operatorname{-0.12131}& \operatorname{-0.224}& \textbf{0.343}\\
0.1829 & \operatorname{-0.0889}&  0.00988& \operatorname{-0.441}& \textbf{0.382}\\
0.0766 & \operatorname{-0.0643}& \operatorname{-0.03049}& \operatorname{-0.114}& \textbf{0.133}\\
\end{array}\right],\; \text{and} \;
\end{equation}
\begin{eqnarray*}
&&\mathbf{a}_0=\left[\operatorname{1.48e-05}, \operatorname{2.51e-05}, \operatorname{2.39e-04}, \operatorname{2.17e-05}, \operatorname{1.50e-05}\right],\\
&&\mathbf{A}=\text{diag}\left[0.0966, 0.0896, 0.0555, 0.1174, 0.1034\right],\\
&&\mathbf{B}=\text{diag}\left[0.8946, 0.8936, 0.6405, 0.8637, 0.8729\right]
\end{eqnarray*}
with the correlation matrix given by
\begin{equation*}
\bR=
\left[
\begin{array}{rrrrr}
1 & 0.7978 & 0.4406 & 0.8036 & 0.6629\\
0.7978 & 1 & 0.5411 & 0.8524 & 0.7998\\
0.4406 & 0.5411 & 1 & 0.5147 & 0.4780\\
0.8036 & 0.8524 & 0.5147 & 1 & 0.7556\\
0.6629 & 0.7998 & 0.4780 & 0.7556 & 1\\
\end{array}\right].
\end{equation*}

It is noted that the last column of the matrix $\bol{\Phi}$ has the largest values. It shows that the influence of the US market on the return indices is larger than those of the domestic ones.

Next, calculate the weights of the three multi-period portfolio strategies (LAMPS, GMV, and partial myopic) as already described in Section 4.1. We exclude the approach of Brandt and Santa Clara in this study since there is no predictable variable within the model (\ref{VAR_int}). We choose $\gamma \in \{5,10,15,20\}$ and $T \in \{4,8,12,16\}$.

\begin{center}
\begin{table}[ptbh]
\label{tab2}
\begin{center}
\begin{tiny}
\scalebox{0.9}{
\begin{tabular}{|c|c|c|c|c|c|}
\hline
\backslashbox{T}{$\gamma$} &5&10&15&20& Method\\ \hline

\multirow{5}{*}{4}

& \textbf{0.5992} (0.0009) &  \textbf{0.5498} (0.0002) & \textbf{0.5335} ($9.4\cdot 10^{-5}$) & \textbf{0.5251} ($5\cdot 10^{-5}$) & LAMPS\\
& 0.5834 (0.0057) &  0.5454 (0.003) & 0.5314 (0.0018) & 0.5237 (0.0012) & GMV\\
& 0.5836 & 0.5456 &  0.5316 &  0.5240 & Part.Myopic\\

\hline

\multirow{5}{*}{8}
& \textbf{0.6002} ($2.6\cdot 10^{-5}$) & \textbf{0.5500} ($6.5\cdot 10^{-6}$)  & \textbf{0.5336} ($3.5\cdot 10^{-6}$)  & \textbf{0.5252} ($2.4\cdot 10^{-6}$) & LAMPS\\
& 0.5833 (0.0083) & 0.5452 (0.0039)  & 0.5309 (0.0023)  & 0.5232 (0.0018) & GMV\\
& 0.5838 & 0.5456 & 0.5316 & 0.5241  & Part.Myopic\\

\hline

\multirow{5}{*}{12}
& \textbf{0.6002} ($2.9\cdot 10^{-6}$) & \textbf{0.5500} ($1.8\cdot 10^{-6}$)  & \textbf{0.5336} ($1.4\cdot 10^{-6}$)  & \textbf{0.5252} ($1.2\cdot 10^{-6}$) &LAMPS\\
& 0.5830 (0.0102) & 0.5448 (0.0045) & 0.5303 (0.0029) & 0.5225 (0.0024) & GMV\\
& 0.5839 & 0.5457 &  0.5316 &  0.5241 & Part.Myopic\\

\hline

\multirow{5}{*}{16}
& \textbf{0.6002} ($1.6\cdot 10^{-6}$)& \textbf{0.5501} ($1.2\cdot 10^{-6}$) & \textbf{0.5336} ($1.12\cdot 10^{-6}$) & \textbf{0.5252} ($10^{-6}$) &LAMPS\\
& 0.5831 (0.0115) & 0.5443 (0.0050)  & 0.5296 (0.0035) & 0.5218 (0.0031) & GMV\\
& 0.5840 & 0.5458 & 0.5317 & 0.5241 & Part.Myopic\\
\hline
\end{tabular}
}
\end{tiny}
\end{center}
\caption{Medians and median absolute deviations (MAD, in parentheses) of the expected quadratic utility for the LAMPS, the multi-period GMV, and the partial myopic strategies in case of weekly MSCI data for the period from the 4th of January, 2002 to the 4th of December, 2009.
%We consider $\gamma \in \{5,10,15,20\}$ and $T\in\{4,8,12,16\}$.
}
\end{table}
\end{center}

\begin{landscape}
\renewcommand{\baselinestretch}{1.0}
\begin{figure}
\caption{\footnotesize The empirical distribution function of the expected quadratic utility for the LAMPS, the multi-period GMV, and the partial myopic in the case of weekly MSCI data for the period from the 4th of January, 2002 to the 4th of December, 2009.
We consider $\gamma \in \{5,10,15,20\}$ and $T=12$.}
\vspace{-0.3cm}
\begin{center}
\includegraphics[angle=270, width=17.8cm]{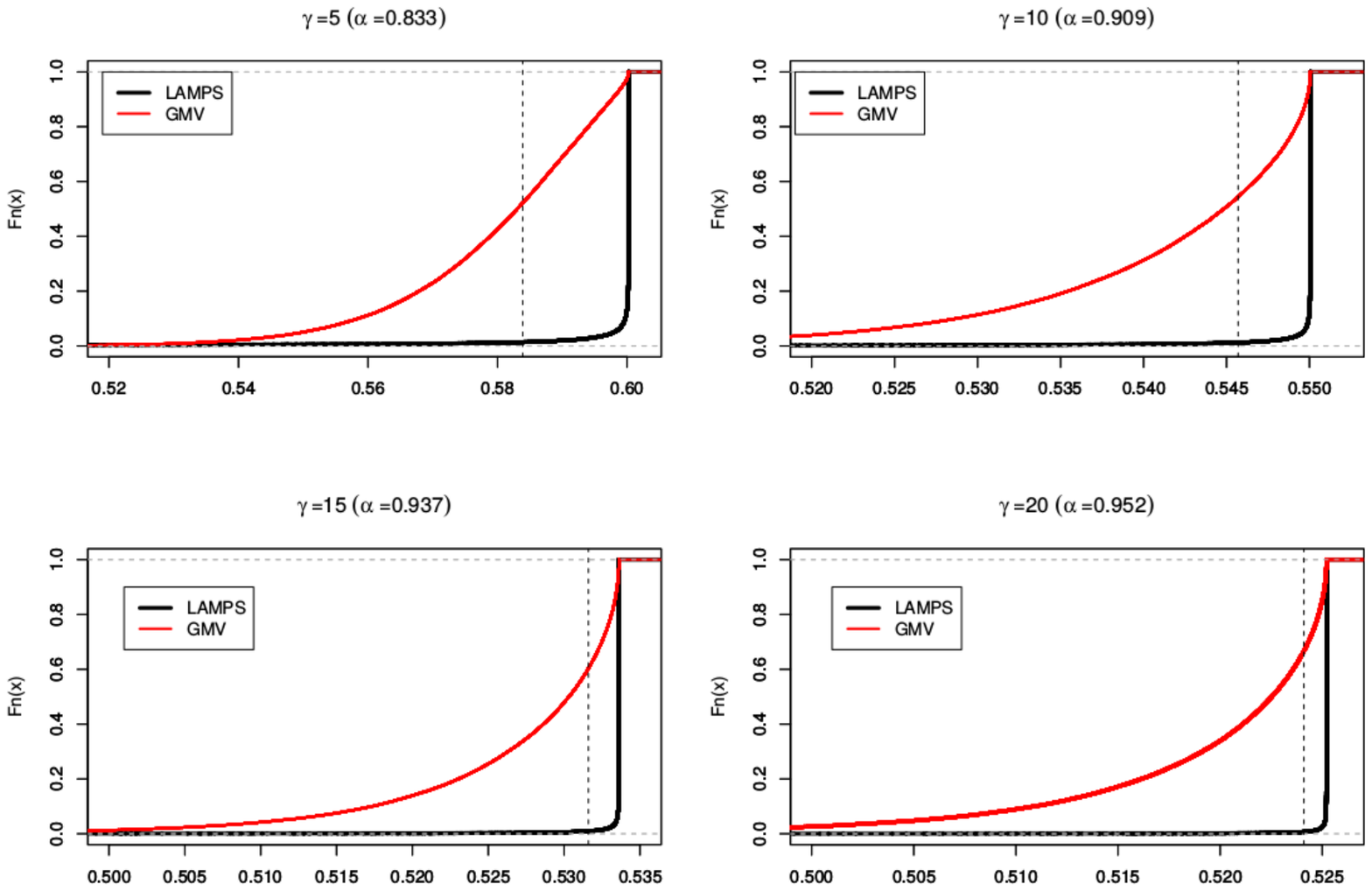}
\end{center}
\label{fig:ECDF_Brandt5}
\end{figure}
\end{landscape}

\begin{landscape}
\renewcommand{\baselinestretch}{1.0}
\begin{figure}
\caption{\footnotesize The median, the $5\%$ and $95\%$ quantiles of the expected quadratic utility for the LAMPS strategy as a function of $T=\{4,8,12,16,20,24,28,32,36,40\}$ in the case of weekly MSCI data (without predictable variable). We consider $\gamma \in \{5,10,15,20\}$.}
\vspace{-0.3cm}
\begin{center}
\includegraphics[width=17.8cm]{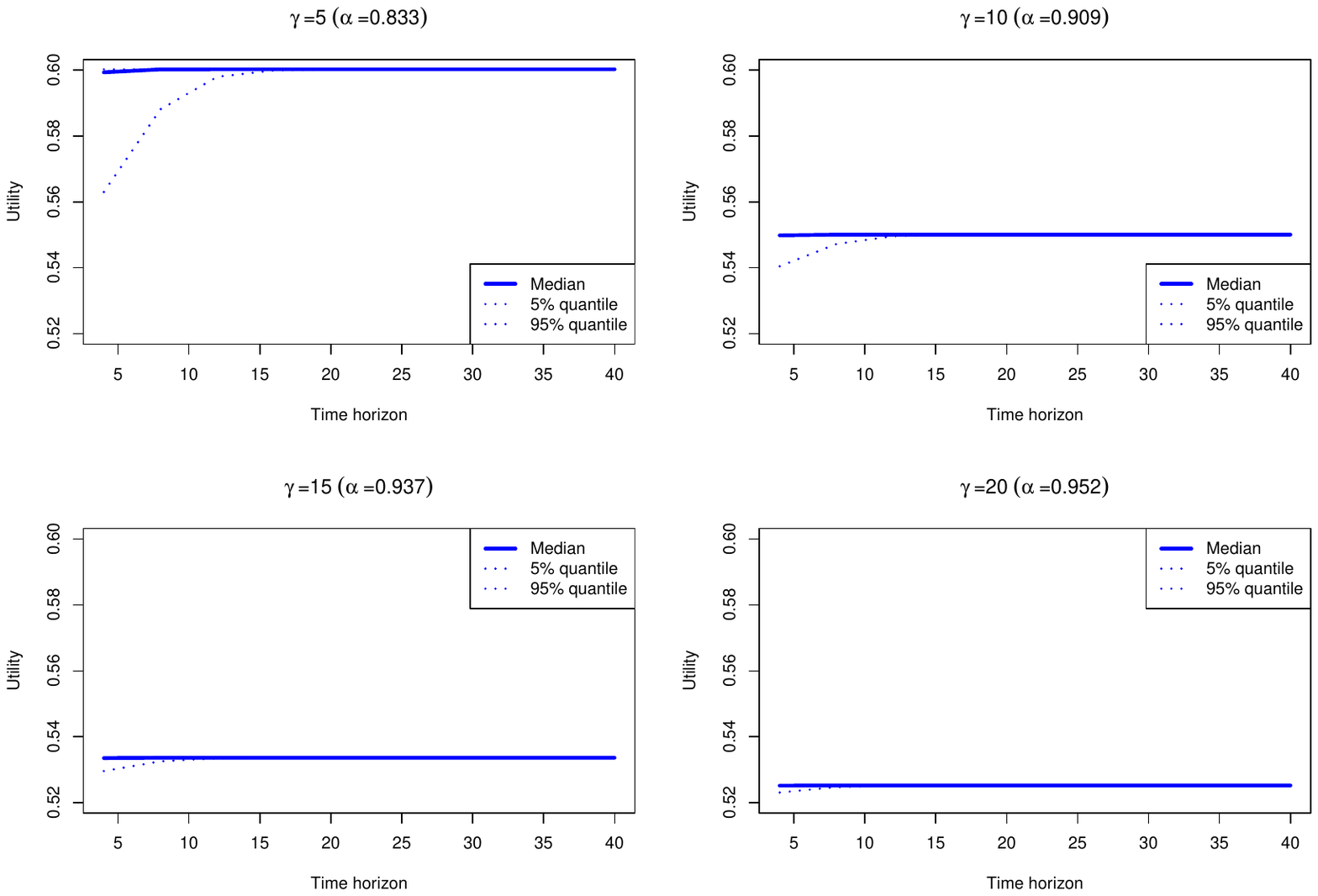}
\end{center}
\label{T1}
\end{figure}
\end{landscape}

In Table 2 we present the medians and the mean absolute deviations for the expected utilities obtained via a simulation study with $10^5$ independent repetitions of the process (\ref{VAR}) with parameters (\ref{VAR_int}).
The results are similar to those of Section 4.1. We observe a very good performance of the LAMPS strategy which turns out to be the best one. On the second place the partial myopic strategy is ranked followed by the multi-period GMV portfolio.

More pronounced results are presented in Figure 6. Here, we observe a very good performance for the LAMPS strategy for all $\gamma$ in the case of $T=12$ weeks. Both  strategies overperform the partial myopic strategy with probability of almost equal to one if $\gamma=5$. For larger values of $\gamma$ a similar behavior is present for the LAMPS approach. The multi-period GMV portfolio performs much worse. For $\gamma=5$ and $\gamma=10$ the probability of getting higher values of the expected utility is about $50\%$, while it is less than $40\%$ for $\gamma=5$ and $\gamma=10$.

Finally, in Figure 7, we plot the median and the corresponding $5\%$ and $95\%$ quantiles calculated for the expected quadratic utility in the case of the LAMPS strategy. In contrast to the example of Section 4.1 we observe that the expected quadratic utility converges faster to its maximum value. A good approximation is already obtained for $T=10$. These results are also in-line with the values presented in Table 2, where the medians of the expected quadratic utility attain their maxima for $T=8$. Furthermore, we observe that the $5\%$ quantile tends to the maximum value for $T \ge 15$ independently of $\gamma$.

Next we want to present a further example. Now the return of the US market index is used as a predictable variable and an investment into four other capital market indexes is considered. This example is motivated by economic theory - the influence of the US market may be larger than of the domestic one, and it is justified in the structure of the matrix $\bol{\Phi}$ (cf. (\ref{VAR_int})). In this example all of the above described strategies are analyzed.

In Table 3 we present the medians and the MADs of the expected utilities. A very good performance of the LAMPS multi-period portfolio strategy is observed. This approach is ranked on the first place. A much worse performance can be observed for the partial myopic strategy, for the multi-period GMV portfolio, and for the BSC method which can be ranked on the second, third, and fourth places.

\begin{center}
\begin{table}[ptbh]

\begin{center}
\begin{tiny}
\scalebox{0.9}{
\begin{tabular}{|c|c|c|c|c|c|}
\hline
\backslashbox{T}{$\gamma$} &5&10&15&20& Method\\ \hline

\multirow{6}{*}{4}
& \textbf{0.6000} (0.0002) &  \textbf{0.5499} ($7.5\cdot10^{-5}$) & \textbf{0.5335} ($7.5\cdot10^{-5}$) & \textbf{0.5251} ($9.1\cdot10^{-5}$) & LAMPS\\

& 0.5835 (0.0078) & 0.5453 (0.0037)  & 0.5310 (0.0022) & 0.5233 (0.0017) & GMV\\

& 0.5836 					& 0.5456 					 &  0.5316 				 &  0.5240 				 & Part.Myopic\\

& 0.4633 (0.1274) &  0.5228 (0.0255) & 0.5236 (0.0094) & 0.5204 (0.0045) & BSC\\
\hline

\multirow{6}{*}{8}

& \textbf{0.6001} (0.0001) & \textbf{0.5499} (0.0001)  & \textbf{0.5335} (0.0001)  & \textbf{0.5251} (0.0001) & LAMPS\\

& 0.5835 (0.0109) & 0.5445 (0.0048)  & 0.5300 (0.0033)  & 0.5221 (0.0028) & GMV\\

& 0.5838 & 0.5456 & 0.5316 & 0.5241  & Part.Myopic\\
&  0.5295 (0.0656) & 0.5320 (0.0168) & 0.5226 (0.0099) & 0.5150 (0.0080) &BSC\\
\hline
\multirow{6}{*}{12}
& \textbf{0.6001} (0.0001) & \textbf{0.5499} (0.0001)  & \textbf{0.5335} (0.0001)  & \textbf{0.5251} (0.0001) &LAMPS\\
& 0.5831 (0.0129) & 0.5435 (0.0058) & 0.5286 (0.0045) & 0.5207 (0.0040) & GMV\\
& 0.5839 & 0.5457 &  0.5316 &  0.5241 & Part.Myopic\\
& 0.5276 (0.0678) & 0.5205 (0.0271) & 0.5055 (0.0216) & 0.4963 (0.0155) &BSC\\
\hline
\multirow{6}{*}{16}
& \textbf{0.6002} ($7.4\cdot 10^{-5}$)& \textbf{0.5500} ($6.8\cdot 10^{-5}$) & \textbf{0.5336} ($6.7\cdot 10^{-5}$) & \textbf{0.5251} ($6.6\cdot 10^{-5}$) &LAMPS\\
& 0.5823 (0.0144)& 0.5422 (0.0071) & 0.5272 (0.0058) & 0.5193 (0.0053) & GMV\\
& 0.5840 & 0.5458 & 0.5317 & 0.5241 & Part.Myopic\\
& 0.5101 (0.0845) & 0.4906 (0.0517) & 0.4730 (0.0372) & 0.4640 (0.0250) &BSC\\
\hline
\end{tabular}
}
\end{tiny}
\end{center}

\caption{Medians and median absolute deviations (MAD, in parentheses) of the expected quadratic utility for the LAMPS, the multi-period GMV, the partial myopic, and the BSC strategies in the case of
weekly MSCI data for the period from the 4th of January, 2002 to the 4th of December, 2009. The investment into the four countries (Belgium, Germany, Japan, and the UK) are considered,
while the returns of the USA market are used as a predictable variable.}
\label{tab3}%
\end{table}
\end{center}

\begin{landscape}
\renewcommand{\baselinestretch}{0.9}
\begin{figure}
\caption{\footnotesize The empirical distribution function of the expected quadratic utility for the LAMPS, the multi-period GMV, and the partial myopic in the case of weekly MSCI data for the period from the 4th of January, 2002 to the 4th of December, 2009. The investment into the four countries (Belgium, Germany, Japan, and the UK) are considered, while the returns of the USA market are used as a predictable variable. ($\gamma \in \{5,10,15,20\}$ and $T=12$).}
\vspace{-0.3cm}
\begin{center}
\includegraphics[angle=270, width=17.8cm]{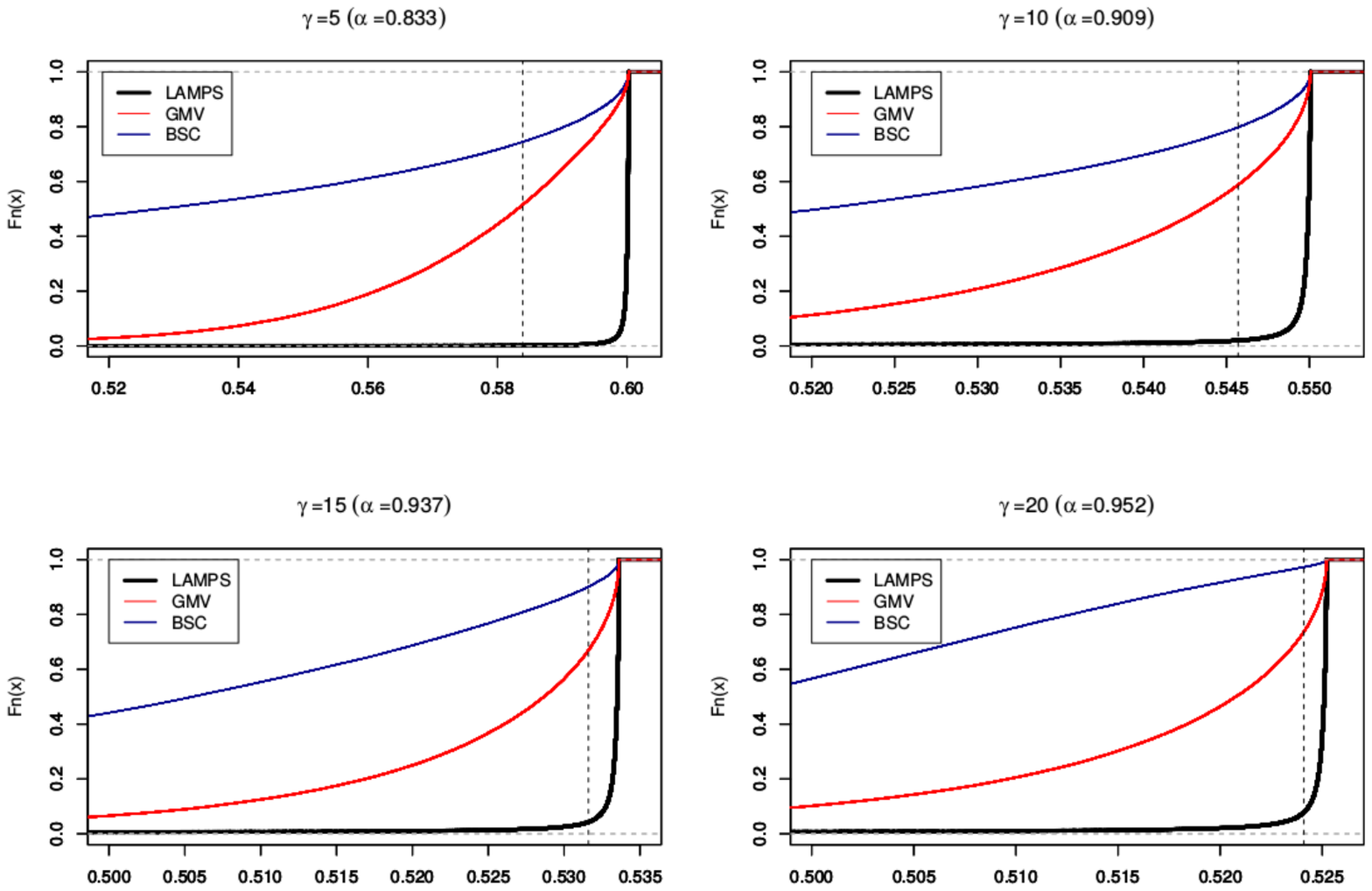}
\end{center}
\label{fig:ECDF_MSCI}
\end{figure}
\end{landscape}

\begin{landscape}
\renewcommand{\baselinestretch}{1.0}
\begin{figure}
\caption{\footnotesize The median, the $5\%$ and $95\%$ quantiles of the expected quadratic utility for the LAMPS strategy as a function of $T=\{4,8,12,16,20,24,28,32,36,40\}$ in the case of weekly MSCI data (with predictable variable).
We consider $\gamma \in \{5,10,15,20\}$.}
\vspace{-0.3cm}
\begin{center}
\includegraphics[width=17.8cm]{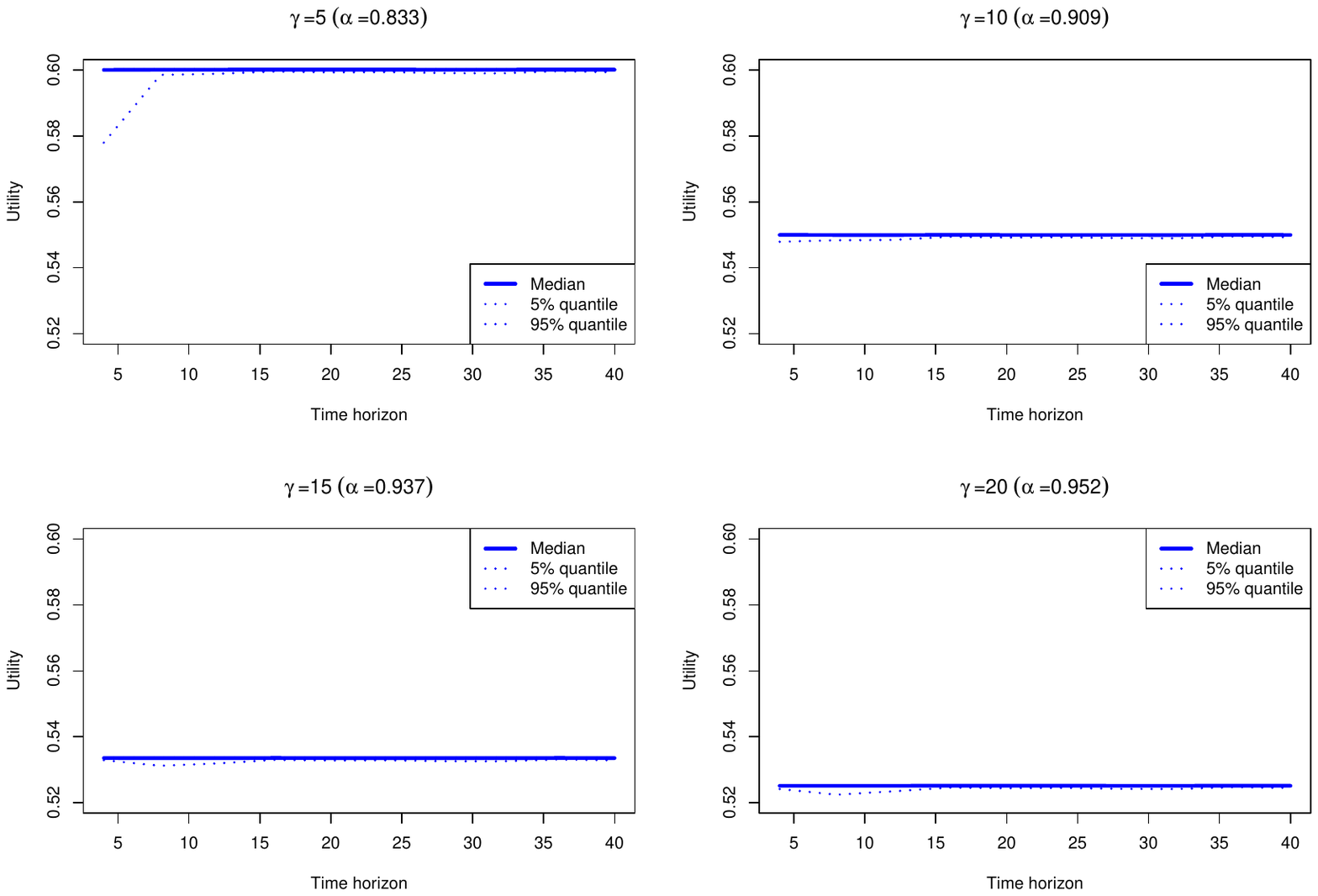}
\end{center}
\label{T2}
\end{figure}
\end{landscape}

The same ranking is also observed in Figure 8 where $T=12$ is used. It has to be emphasized that the LAMPS strategy overperforms the partial myopic strategy with probability almost $1$ for all of the considered values of $\gamma$, while the multi-period GMV portfolio and the BSC method are really bad. Figure 9 shows that the expected quadratic utility converges to its maximum value fast, since the median coincides with the $95\%$ quantile and the maximum values already for small values of $T\le 8$.

The results of our empirical studies lead to the following conclusive remarks:
\begin{itemize}
 \item  The proposed method (LAMPS) shows superior performance for both VAR(1) and VAR(1)-CCC-GARCH(1,1) processes estimated from the data. It can be applied in the cases with and without predictable variables. Furthermore, a considerable improvement in terms of the expected utility is obtained if the time horizon and/or the re-balancing frequency of the portfolio weights increase.

 \item The LAMPS strategy achieves the maximum utility gain when the time horizon $T$ increases. Indeed, the quadratic utility function (\ref{QUF}) attains its maximum\footnote{It can be easily shown that taking the derivative of $W_t-\dfrac{\alpha}{2}W_t^2$ with respect to $W_t$ and setting it equal to zero leads to the maximum attainable wealth level at time point $t$ which is equal to $1/\alpha$.} at the point $W_t=1/\alpha$, which leads to the maximum utility gain equal to $1/(2\alpha)\in\{0.6002, 0.5501, 0.5336, 0.5252\}$ if $\alpha\in\{0.833, 0.909, 0.937, 0.952\}$, respectively. Tables 2 and 3 show that even for $T$ at around $8$ this level is achieved with a small deviation.

 \item The inclusion of the predictable variable into the model leads to interesting results. On one side, the median converges faster to the maximum value in case of small $\gamma$. On the other side, a number of outliers is present which move the $5\%$ quantile away from the maximum value.

 \item  It has to be mentioned that the influence of the re-balancing frequency of the portfolio weights on the performance of the LAMPS strategy is remarkable. In general we observe a better performance for weekly data. For monthly data LAMPS has a little bit more uncertainty in the expected utility gains. This fact shows that the increase of the re-balancing frequency improves the proposed method.
\end{itemize}

The above mentioned observations indicate that the suggested LAMPS multi-period portfolio strategy is highly recommended for long time horizons and/or high re-balancing frequencies which is exactly the case when the approximate method of Brandt and Santa Clara (2006) deviates from the exact one significantly.

\section{Summary}

\noindent Although in 1959 Markowitz has already formulated the multi-period asset allocation problem, no closed-form solution is available in literature up to now. Brandt and Santa Clara (2006) provided an analytical solution by imposing some constraints on the structure of the portfolio weights and, thus, by transforming the multi-period portfolio selection problem with time-dependent weights into a more simpler one where the matrix of time-independent constants has to be calculated at the beginning of the investment horizon. Moreover, note that the approach may lead to a local maximum and not a global one.

In the present paper, we derive a closed-form solution of the multi-period portfolio selection problem with and without a riskless asset which is obtained under weak conditions on the process of the asset returns. The expressions of the optimal portfolio weights are based only on the conditional mean vectors and the conditional covariance matrices. Moreover, under the additional assumption of independence they are similar to the solutions of the single-period optimization problems that are performed at each time point. We prove that only the coefficient of the shape parameter depends on the dynamics of the asset returns in the case without a riskless asset. In the presence of a riskless asset the solutions of the multi-period problem and of the sequence of simple-period problems are proportional where only the coefficients of proportionality depend on the process dynamics. A very remarkable result is obtained for the portfolio selection problem based on the tangency portfolio. Assuming independent returns it is proved that the solution of the multi-period problem and the solutions of the simple-period problems are the same.

The derived multi-period portfolio choice strategies are compared with the existing methods proposed in literature for two real data sets. The first data are taken from the paper of Brandt and Santa Clara (2006), while the returns of five developed stock market indices are used in the second example. We observe a very good performance of the predictive loss approximation of the multi-period strategy in case with a riskless asset which is always ranked on the first place.

The obtained results can be further extended by allowing the intermediate consumptions. This can be done without any large effort by slightly modifying the weights of Theorems 1, 2 and 3. Another possibility to extend the obtained results is to impose some constraints on the structure of the portfolio weights. It is remarkable that if the considered short constraints are based on linear restrictions imposed on the portfolio weights then the multi-period portfolio selection problem can be treated in the same way. This leads to an interesting extension of the results obtained in our paper. The inequality constraint would require to develop additional theoretical results which will be treated in a separate paper.

\section{Appendix}
\noindent In this section the proofs of the theorems are given.\\[0.1cm]

\noindent \textbf{Proof of Theorem 1:}

\noindent First, we note that the expression of the optimal weights at period $T-1$ is given in (\ref{weights_T-1}). The rest of the proof is done by using the mathematical induction on the expressions of the portfolio weights and the value function.
Let
\begin{equation}\label{aaa}
\bA_{i}=E_{i-1}[V_{i+1}\tbx_{i}\tbx^\prime_{i}]~~~~\text{for}~~~~i=1,\ldots,T-1~~~~\text{and}~~~~A_T=\bSigma_T+\tbm_T\tbm_T^\prime\,.
\end{equation}
Moreover, let
\begin{equation}
\tbm^{*}_{i}= \left\{
  \begin{array}{l l}
    \tbm_{T} &\quad\text{for}~~~~i=T\\
    E_{i-1}[R_{i+1}\tbx_{i}]& \quad \text{for}~~~ i=1,\ldots,T-1\,,
  \end{array} \right.
\end{equation}
$R_{i}=\frac{\bii^{\prime}\bA_{i}^{-1}\tbm^*_i}{\bii^{\prime}\bA^{-1}_i\bii}$, $V_{i}=\frac{1}{\bii^{\prime}\bA^{-1}_i\bii}$ and $\tilde{s}_i=\tbm_i^{*\;\prime} \tilde{\bQ}_i\tbm^*_i$ with
\[\tilde{\bQ}_i=\bA^{-1}_{i}-
\frac{\bA^{-1}_{i}\bii\bii^{\prime}\bA^{-1}_{i}}{\bii^{\prime}\bA^{-1}_{i}\bii} \quad \text{for} \quad i=1,...,T-1 \,.\]
Note, that
\[\bii^{\prime}\tilde{\bQ}_{i}=\bzero^{\prime},~~~~\tilde{\bQ}_{i}\bii=\bzero,~~~~ \tilde{\bQ}_{i}\bA_{i}\tilde{\bQ}_{i}=\tilde{\bQ}_{i}\,.\]

Let $\bw^{*\;\prime}_{T-1}$ be the optimal portfolio weights calculated for period $T-1$ given in (\ref{weights_T-1}).
First, we calculate the value function for period $T-2$. It holds that
{\small
\begin{eqnarray*}
&&V(T-2,W_{T-2},\mathcal{F}_{T-2})\\
&=&\max\limits_{\bw_{T-2}:\bw^{\prime}_{T-2}\bii=1}E_{T-2}\Big{[}\max\limits_{\bw_{T-1}:\bw^{\prime}_{T-1}\bii=1}\left(W_{T-1}\bw^{\prime}_{T-1}\tbm_{T}
-\frac{\alpha}{2}W^2_{T-1}\bw^{\prime}_{T-1}\bA_{T}\bw_{T-1}\right)\Big{]}\\
&=&\max\limits_{\bw_{T-2}:\bw^{\prime}_{T-2}\bii=1}E_{T-2}\Big{[}W_{T-1}\bw^{*\;\prime}_{T-1}\tbm_{T}-\frac{\alpha}{2}W^2_{T-1}\bw^{*\;\prime}_{T-1}\bA_{T}\bw^*_{T-1}\Big{]}\\
&=&\max\limits_{\bw_{T-2}:\bw^{\prime}_{T-2}\bii=1}E_{T-2}\Big{[}W_{T-1}\left(\frac{\bA^{-1}_T\bii}{\bii^{\prime}\bA^{-1}_T\bii}+\frac{1}{\alpha W_{T-1}}\tilde{\bQ}_T\tbm_T\right)^{\prime}\tbm_T\\
&-&\frac{\alpha}{2}W^2_{T-1}\left(\frac{\bA^{-1}_T\bii}{\bii^{\prime}\bA^{-1}_T\bii}+\frac{1}{\alpha W_{T-1}}\tilde{\bQ}_T\tbm_T\right)^{\prime}\bA_T\left(\frac{\bA^{-1}_T\bii}{\bii^{\prime}\bA^{-1}_T\bii}+\frac{1}{\alpha W_{T-1}}\tilde{\bQ}_T\tbm_T\right)\Big{]}\\
&=&\max\limits_{\bw_{T-2}:\bw^{\prime}_{T-2}\bii=1}E_{T-2}\Big{[}W_{T-1}\left(\frac{\bii^{\prime}\bA_{T}^{-1}\tbm_T}{\bii^{\prime}\bA^{-1}_{T}\bii}
-\frac{1}{2}\underbrace{\frac{\tbm_T^{\prime}\tilde{\bQ}_T\bA_T\bA_{T}^{-1}\bii}{\bii^{\prime}\bA^{-1}_{T}\bii}}_{=0}
-\frac{1}{2}\underbrace{\frac{\bii^{\prime}\bA_{T}^{-1}\bA_{T}\tilde{\bQ}_T\tbm_T}{\bii^{\prime}\bA^{-1}_{T}\bii}}_{=0}\right)\\
&+&\frac{1}{2\alpha}\tbm_T^{\prime}\tilde{\bQ}_T\tbm_T-\frac{\alpha}{2}\frac{W^2_{T-1}}{\bii^{\prime}\bA^{-1}_T\bii}\Big{]}\,,
\end{eqnarray*}
}

Using the definitions of $R_i$, $V_i$ and $\tilde{s}_i$ we obtain
\begin{eqnarray*}
&&V(T-2,W_{T-2},\mathcal{F}_{T-2})\\
&=&\max\limits_{\bw_{T-2}:\bw^{\prime}_{T-2}\bii=1}E_{T-2}\Big{[}W_{T-1}R_T+\frac{1}{2\alpha}\tilde{s}_T-\frac{\alpha}{2}W^2_{T-1}V_T\Big{]}\\
&=&\max\limits_{\bw_{T-2}:\bw^{\prime}_{T-2}\bii=1}E_{T-2}\Big{[}W_{T-2}\bw^{\prime}_{T-2}R_T\tbx_{T-1}+\frac{1}{2\alpha}\tilde{s}_T
-\frac{\alpha}{2}W^2_{T-2}V_T(\bw^{\prime}_{T-2}\tbx_{T-1})^2\Big{]}\\
%&=&\max\limits_{\bw_{T-2}:\bw^{\prime}_{T-2}\bii=1}\Big{[}W_{T-2}R_{GMV,T}\bw^{\prime}_{T-2}\tbm_{T-1}+\frac{1}{2\alpha}s_T
%-\frac{\alpha}{2}W^2_{T-2}V_{GMV,T}\left(\bw^{\prime}_{T-2}\bSigma_{T-1}\bw_{T-2}+(\bw^{\prime}_{T-2}\tbm_{T-1})^2\right)\Big{]}\\
&=&\max\limits_{\bw_{T-2}:\bw^{\prime}_{T-2}\bii=1}\Big{[}W_{T-2}\bw^{\prime}_{T-2}\tbm^*_{T-1}+F(\tilde{s}_T)
-\frac{\alpha}{2}W^2_{T-2}\left(\bw^{\prime}_{T-2}\bA_{T-1}\bw_{T-2}\right)\Big{]}\,,
\end{eqnarray*}
where
\begin{equation}\label{FsT}
F(\tilde{s}_T)=\frac{1}{2\alpha}E_{T-2}[\tilde{s}_T]\,.
\end{equation}
$F(\tilde{s}_T)$ does not depend on $\bw_{T-2}$.

The last expression is similar to the value function at period $T-1$ (cf. (\ref{Vt_1})). Hence, the optimal weights $\bw_{T-2}^*$ are given by
\begin{equation}\label{weights_T-2}
\bw^*_{T-2}=\frac{\bA^{-1}_{T-1}\bii}{\bii^{\prime}\bA^{-1}_{T-1}\bii}+\frac{1}{\alpha W_{T-2}}\tilde{\bQ}_{T-1}\tbm^*_{T-1}~~\text{with}~~\tilde{\bQ}_{T-1}=\bA^{-1}_{T-1}-\frac{\bA^{-1}_{T-1}\bii\bii^{\prime}\bA^{-1}_{T-1}}{\bii^{\prime}\bA^{-1}_{T-1}\bii}\,.
\end{equation}
As a result, the following expressions are the basis of the induction
\begin{eqnarray*}
V(T-2,W_{T-2},\mathcal{F}_{T-2})
&=&\max\limits_{\bw_{T-2}:\bw^{\prime}_{T-2}\bii=1}\Big{[}W_{T-2}\bw^{\prime}_{T-2}\tbm^*_{T-1}+F(\tilde{s}_T)\\
&-&\frac{\alpha}{2}W^2_{T-2}\bw^{\prime}_{T-2}\bA_{T-1}\bw_{T-2}\Big{]}\\
\bw^*_{T-2}&=&\frac{\bA^{-1}_{T-1}\bii}{\bii^{\prime}\bA^{-1}_{T-1}\bii}+\frac{1}{\alpha W_{T-2}}\bQ_{T-1}\tbm^*_{T-1}
\end{eqnarray*}
with $F(\tilde{s}_T)$ as defined in (\ref{FsT}).

%==================================================================================================================================================================================================

In the induction hypothesis we assume that the statement holds for $t=n$, i.e.,
\begin{eqnarray*}
V(T-n,W_{T-n},\mathcal{F}_{T-n})&=&\max\limits_{\bw_{T-n}:\bw^{\prime}_{T-n}\bii=1}\Big{[}W_{T-n}\bw^{\prime}_{T-n}\tbm^*_{T-n+1}
\\
&&\hspace{-4cm}+F(\tilde{s}_T,\tilde{s}_{T-1},\ldots,\tilde{s}_{T-n+2})-\frac{\alpha}{2}W^2_{T-n}\bw^{\prime}_{T-n}\bA_{T-n+1}\bw_{T-n}\Big{]}
\\
\bw^*_{T-n}&=&\frac{\bA^{-1}_{T-n+1}\bii}{\bii^{\prime}\bA^{-1}_{T-n+1}\bii}+\frac{1}{\alpha W_{T-n}}\tilde{\bQ}_{T-n+1}\tbm^*_{T-n+1}\,,
\end{eqnarray*}
where
\begin{equation*}
F(\tilde{s}_T,\tilde{s}_{T-1},\ldots,\tilde{s}_{T-n+2})
=\frac{1}{2\alpha}\left(E_{T-2}[\tilde{s}_T]+\sum\limits_{m=T-n+2}^{T-1}E_{m-2}[\tilde{s}_m]\right)\,.
\end{equation*}
Note that the last quantity does not depend on $\bw_{T-n}$.

%%%%%%%%%%%%%%%%%%%%%%%%%%%%%%%%%%%%%%%%%%%%%%%%%%%%%%%%%%%%%%%%%%%%%%%%%%%%%%%%%%%%%%%%%%%%%%%%%
In the inductive step we prove that the last identities also hold for $t=n+1$. It is sufficient to derive the value function for period $T-(n+1)$ which is given by
{\scriptsize
\begin{eqnarray*}
&&V(T-(n+1),W_{T-(n+1)},\mathcal{F}_{T-(n+1)})\\
&=&\max\limits_{\bw^{\prime}_{T-(n+1)}\bii=1}E_{T-(n+1)}\left(W_{T-n}\bw^{*\;\prime}_{T-n}\tbm^*_{T-n+1}+F(\tilde{s}_T,\tilde{s}_{T-1},\ldots,\tilde{s}_{T-n+2})\right.\\
&-&\left.\frac{\alpha}{2}W^2_{T-n}\bw^{*\;\prime}_{T-n}\bA_{T-n+1}\bw^*_{T-n}\right)\\
&=&\max\limits_{\bw^{\prime}_{T-(n+1)}\bii=1}E_{T-(n+1)}\Big{[}W_{T-n}\left(\underbrace{\frac{\bii^{\prime}\bA^{-1}_{T-n+1}\tbm^*_{T-n+1}}{\bii^{\prime}\bA^{-1}_{T-n+1}\bii}}_{=R_{T-n+1}}
-\frac{1}{2}\underbrace{\frac{\tbm_{T-n+1}^{*\;\prime}\tilde{\bQ}_{T-n+1}\bii}{\bii^{\prime}\bA^{-1}_{T-n+1}\bii}}_{=0}
-\frac{1}{2}\underbrace{\frac{\bii^{\prime}\tilde{\bQ}_{T-n+1}\tbm^*_{T-n+1}}{\bii^{\prime}\bA^{-1}_{T-n+1}\bii}}_{=0}\right)\\
&+&F(\tilde{s}_T,\tilde{s}_{T-1},\ldots,\tilde{s}_{T-n+2})+\frac{1}{2\alpha}\underbrace{\tbm^{*\;\prime}_{T-n+1}\tilde{\bQ}_{T-n+1}\tbm^*_{T-n+1}}_{=\tilde{s}_{T-n+1}}
-\frac{\alpha}{2}\underbrace{\frac{1}{\bii^{\prime}\bA^{-1}_{T-n+1}\bii}}_{=V_{T-n+1}}W^2_{T-n}\Big{]}\,.
\end{eqnarray*}
}
Hence, we obtain
\begin{eqnarray*}
&&V(T-(n+1),W_{T-(n+1)},\mathcal{F}_{T-(n+1)})\\
&=&\max\limits_{\bw^{\prime}_{T-(n+1)}\bii=1}E_{T-(n+1)}\Big{[}W_{T-(n+1)}\bw^{\prime}_{T-(n+1)}\tbx_{T-n}R_{T-n+1}\\
&+&F(\tilde{s}_T,\tilde{s}_{T-1},\ldots,\tilde{s}_{T-n+1})-\frac{\alpha}{2}W^2_{T-(n+1)}V_{T-n+1}(\bw^{\prime}_{T-n+1}\tbx_{T-n})^2\Big{]}\\
&=&\max\limits_{\bw^{\prime}_{T-(n+1)}\bii=1}\left(W_{T-(n+1)}\bw^{\prime}_{T-(n+1)}\tbm^*_{T-n}+F(\tilde{s}_T,\tilde{s}_{T-1},\ldots,\tilde{s}_{T-n+1})\right.\\
&-&\left.\frac{\alpha}{2}W^2_{T-(n+1)}\bw^{\prime}_{T-(n+1)}\bA_{T-n}\bw_{T-(n+1)}\right)\,,
\end{eqnarray*}
where
\[F(\tilde{s}_T,\tilde{s}_{T-1},\ldots,\tilde{s}_{T-n+1})=F(s_T,\tilde{s}_{T-1},\ldots,\tilde{s}_{T-n+2})+\frac{1}{2\alpha}E_{T-(n+1)}[\tilde{s}_{T-n+1}]\,.\]

It is the desired form of the value function at period $T-(n+1)$. Because this expression is similar to the value function at period $T-n$, we get the following formula for the weights at period $T-(n+1)$
\begin{equation*}
\bw^*_{T-(n+1)}=\frac{\bA^{-1}_{T-n}\bii}{\bii^{\prime}\bA^{-1}_{T-n}\bii}+\frac{1}{\alpha W_{T-(n+1)}}\tilde{\bQ}_{T-n}\tbm^*_{T-n}
\,.
\end{equation*}

The theorem is proved. \\[0.3cm]

For proving Corollary 1 we use the result of Proposition 6.1.

\begin{proposition}
Let $\bx$ be a random vector with mean $\bmu$ and positive definite covariance matrix $\bSigma$. Let $\bA=\bSigma+\tbm\tbm^\prime$ with $\tbm=\bmu+\bii$. If
\begin{equation} \label{wA}
\bw=\frac{\bA^{-1}\bii}{\bii^{\prime}\bA^{-1}\bii}+\tilde{\alpha}^{-1}\tilde{\bQ}\tilde{\bmu}~~~~\text{with}~~~~\tilde{\bQ}=\bA^{-1}-\frac{\bA^{-1}\bii\bii^\prime\bA^{-1}}{\bii^\prime\bA^{-1}\bii}
\end{equation}
then
\begin{equation} \label{wE}
\bw=\frac{\bSigma^{-1}\bii}{\bii^{\prime}\bSigma^{-1}\bii}+\alpha^{-1}\bQ\tilde{\bmu}~~~~\text{with}~~~~\bQ=\bSigma^{-1}-\frac{\bSigma^{-1}\bii\bii^\prime\bSigma^{-1}}{\bii^\prime\bSigma^{-1}\bii}
\end{equation}
and
\begin{equation} \label{alpha}
\alpha^{-1}=\frac{\tilde{\alpha}^{-1}(\bii^\prime\bSigma^{-1}\bii)-\bii^\prime\bSigma^{-1}\tbm}{(1+\tbm^\prime\bSigma^{-1}\tbm)\bii^\prime\bSigma^{-1}\bii-(\bii^\prime\bSigma^{-1}\tbm)^2}=
\frac{\tilde{\alpha}^{-1}-1-R_{GMV}}{1+s}\,,
\end{equation}
where $R_{GMV}=\frac{\bii^\prime\bSigma^{-1}\bmu}{\bii^\prime\bSigma^{-1}\bii}, ~~~~s=\tbm^{\prime}\bQ\tbm=\bmu^{\prime}\bQ\bmu$.
\end{proposition}

\noindent \textbf{Proof of Proposition 1:}

\noindent From (\ref{wE}) we obtain
\begin{equation} \label{wE1}
\bw=\left(\frac{1}{\bii^\prime\bSigma^{-1}\bii}-\alpha^{-1}\frac{\bii^\prime\bSigma^{-1}\tbm}{\bii^\prime\bSigma^{-1}\bii}\right)\bSigma^{-1}\bii+\alpha^{-1}\bSigma^{-1}\tbm
=C_1\bSigma^{-1}\bii+C_2\bSigma^{-1}\tbm\,,
\end{equation}
where
\begin{equation}\label{qw}
C_1=\frac{1}{\bii^{\prime}\bSigma^{-1}\bii}-C_2\frac{\bii^{\prime}\bSigma^{-1}\tbm}{\bii^{\prime}\bSigma^{-1}\bii} \quad \text{and} \quad C_2=\alpha^{-1} \,.
\end{equation}

In order to prove the proposition we need to show that (\ref{wA}) can be expressed in the same way. The application of the Sherman-Morrison formula (Harville (1997, Theorem 18.2.8)), i.e.,
$$\bA^{-1}=(\bSigma+\tbm\tbm^{\prime})^{-1}=\bSigma^{-1}-\frac{\bSigma^{-1}\tbm\tbm^{\prime}\bSigma^{-1}}{1+\tbm^\prime\bSigma^{-1}\tbm}$$
leads to
\begin{eqnarray}\label{wE2}
\bw&=&(1-K\tilde{\alpha}^{-1})\frac{1+\tbm^\prime\bSigma^{-1}\tbm}{(1+\tbm^\prime\bSigma^{-1}\tbm)\bii^\prime\bSigma^{-1}\bii-(\bii^\prime\bSigma^{-1}\tbm)^2}\bSigma^{-1}\bii\\
&+&\left(-\frac{\tbm^\prime\bSigma^{-1}\bii(1-K\tilde{\alpha}^{-1})}{(1+\tbm^\prime\bSigma^{-1}\tbm)\bii^\prime\bSigma^{-1}\bii-(\bii^\prime\bSigma^{-1}\tbm)^2}
+\frac{\tilde{\alpha}^{-1}}{1+\tbm^\prime\bSigma^{-1}\tbm}\right)\bSigma^{-1}\tbm \,, \nonumber
%&=&D_1\bSigma^{-1}\bii+D_2\bSigma^{-1}\tbm\,,
\end{eqnarray}
where
\begin{equation}\label{K}
K=\bii^\prime\bA^{-1}\tbm=\frac{\bii^\prime\bSigma^{-1}\tbm}{1+\tbm^\prime\bSigma^{-1}\tbm}.
\end{equation}

From the structure of (\ref{wE1}) and (\ref{wE2}) we get
\begin{eqnarray*}
\alpha^{-1}&=&C_2=\left(-\frac{\tbm^\prime\bSigma^{-1}\bii(1-K\tilde{\alpha}^{-1})}{(1+\tbm^\prime\bSigma^{-1}\tbm)\bii^\prime\bSigma^{-1}\bii-(\bii^\prime\bSigma^{-1}\tbm)^2}\right)
+\frac{\tilde{\alpha}^{-1}}{1+\tbm^\prime\bSigma^{-1}\tbm}
\\
&=&\frac{\tilde{\alpha}^{-1}(\bii^\prime\bSigma^{-1}\bii)-\bii^\prime\bSigma^{-1}\tbm}{(1+\tbm^\prime\bSigma^{-1}\tbm)
\bii^\prime\bSigma^{-1}\bii-(\bii^\prime\bSigma^{-1}\tbm)^2}=\frac{\tilde{\alpha}^{-1}-1-R_{GMV}}{1+s}\,.
\end{eqnarray*}

For proving the proposition we only need to show the equality of the coefficients in front of $\bSigma^{-1}\bii$ in (\ref{wE1}) and (\ref{wE2}). It holds that
{\small
\begin{eqnarray*}
C_1&=&\frac{1}{\bii^{\prime}\bSigma^{-1}\bii}-C_2\frac{\bii^{\prime}\bSigma^{-1}\tbm}{\bii^{\prime}\bSigma^{-1}\bii}=
\frac{1}{\bii^\prime\bSigma^{-1}\bii}\\ &-&\left(\frac{\tilde{\alpha}^{-1}}{1+\tbm^\prime\bSigma^{-1}\tbm}-\frac{\tbm^\prime\bSigma^{-1}\bii(1-K\tilde{\alpha}^{-1})}{(1+\tbm^\prime\bSigma^{-1}\tbm)\bii^\prime\bSigma^{-1}\bii-(\bii^\prime\bSigma^{-1}\tbm)^2}\right)\frac{\bii^\prime\bSigma^{-1}\tbm}{\bii^\prime\bSigma^{-1}\bii}\\
&=&\frac{(1-K\tilde{\alpha}^{-1})}{\bii^\prime\bSigma^{-1}\bii}+\frac{(\bii^\prime\bSigma^{-1}\tbm)^2(1-K\tilde{\alpha}^{-1})}{\bii^\prime\bSigma^{-1}\bii\left((1+\tbm^\prime\bSigma^{-1}\tbm)\bii^\prime\bSigma^{-1}\bii-(\bii^\prime\bSigma^{-1}\tbm)^2\right)}\\
&=&\frac{(1-K\tilde{\alpha}^{-1})}{\bii^\prime\bSigma^{-1}\bii}\left(\frac{(1+\tbm^\prime\bSigma^{-1}\tbm)\bii^\prime\bSigma^{-1}\bii}{(1+\tbm^\prime\bSigma^{-1}\tbm)\bii^\prime\bSigma^{-1}\bii-(\bii^\prime\bSigma^{-1}\tbm)^2}\right)\,.
\end{eqnarray*}
}

The last identity completes the proof.\\[0.3cm]

\noindent \textbf{Proof of Corollary 1:}

\noindent Under the assumption of independence it holds that
\begin{equation}\label{bA_in}
\bA_{T-t+1}= \left\{
  \begin{array}{l l}
    \bSigma_T+\tbm_T\tbm^\prime_T &\quad\text{for}~~~~t=1\\
    V_{T-t+2}(\bSigma_{T-t+1}+\tbm_{T-t+1}\tbm^\prime_{T-t+1}) & \quad \text{for}~~~ t=2,\ldots,T\,,
  \end{array} \right.
\end{equation}
and
\begin{equation}\label{mu_in}
\tbm^*_{T-t+1}= \left\{
  \begin{array}{l l}
    \tbm_T &\quad\text{for}~~~~t=1\\
    R_{T-t+2} \tbm_{T-t+1} & \quad \text{for}~~~ t=2,\ldots,T\\
  \end{array} \right. \,.
\end{equation}

Let $R_{T+1}=V_{T+1}=1$. Then,
{\small
\begin{eqnarray*}
\bw^*_{T-t}&=&\frac{(V_{T-t+2}(\bSigma_{T-t+1}+\tbm_{T-t+1}\tbm^\prime))^{-1}\bii}{\bii^{\prime}(V_{T-t+2}(\bSigma_{T-t+1}+\tbm_{T-t+1}\tbm^\prime_{T-t+1}))^{-1}\bii}
+\frac{1}{\alpha W_{T-t}}\tilde{\bQ}_{T-t+1} (R_{T-t+2}\tbm_{T-t+1})\\
&=&
\frac{(\bSigma_{T-t+1}+\tbm_{T-t+1}\tbm^\prime_{T-t+1})^{-1}\bii}{\bii^{\prime}(\bSigma_{T-t+1}+\tbm_{T-t+1}\tbm^\prime_{T-t+1})^{-1}\bii}
+\frac{R_{T-t+2}}{\alpha W_{T-t}V_{T-t+2}}\tilde{\tilde{\bQ}}_{T-t+1} \tbm_{T-t+1}
\end{eqnarray*}
}
with
{\small
\begin{eqnarray*}
\tilde{\tilde{\bQ}}_{T-t+1}&=&(\bSigma_{T-t+1}+\tbm_{T-t+1}\tbm^\prime_{T-t+1})^{-1}\\
&-&\frac{(\bSigma_{T-t+1}+\tbm_{T-t+1}\tbm^\prime_{T-t+1})^{-1}\bii
\bii' (\bSigma_{T-t+1}+\tilde{\bmu}_{T-t+1}\tilde{\bmu}_{T-t+1}^\prime )^{-1}}{\bii^\prime(\bSigma_{T-t+1}+\tbm_{T-t+1}\tbm^\prime_{T-t+1})^{-1}\bii}
%
%\frac{(\bSigma_{T-t+1}+\tbm_{T-t+1}\tbm^\prime_{T-t+1})^{-1}\bii \bii^\prime (\bSigma_{T-t+1}+\tbm_{T-t+1}\tbm^\prime_{T-t+1})^{-1}}{\bii^\prime (\bSigma_{T-t+1}+\tbm_{T-t+1}\tbm^\prime_{T-t+1})^{-1} \bii}
\end{eqnarray*}
}
and
\[\frac{R_{T-t+2}}{V_{T-t+2}}= \prod\limits_{i=T-t+2}^{T}\frac{\frac{\bii^{\prime}(\bSigma_i+\tbm_i\tbm_i^\prime)^{-1}\tbm_i}{\bii^{\prime}(\bSigma_i+\tbm_i\tbm_i^\prime)^{-1}\bii}}
{\frac{1}{\bii^{\prime}(\bSigma_i+\tbm_i\tbm_i^\prime)^{-1}\bii}} = \prod\limits_{i=T-t+2}^{T} \bii^{\prime}(\bSigma_i+\tbm_i\tbm_i^\prime)^{-1}\tbm_i \,,\]
where the last identity follows from the definition of $R_{T-t+2}$ and $V_{T-t+2}$ given in (\ref{R}).

The rest of the proof follows from Proposition 1 if $\bSigma$ is replaced by $\bSigma_{T-t+1}$, $\tbm$ by $\tbm_{T-t+1}$  and
{\footnotesize
\[\tilde{\alpha}^{-1}=\frac{1}{\alpha W_{T-t}}\frac{R_{T-t+2}}{V_{T-t+2}}=  \frac{1}{\alpha W_{T-t}}\left(\prod\limits_{i=T-t+2}^{T}\bii^{\prime}(\bSigma_i+\tbm_i\tbm_i^\prime)^{-1}\tbm_i\right)=\frac{1}{\alpha W_{T-t}}\left(\prod\limits_{i=T-t+2}^{T}a_i\right)\,,\]
}
where
\[a_i=\bii^{\prime}(\bSigma_i+\tbm_i\tbm_i^\prime)^{-1}\tbm_i=\frac{1+R_{GMV,i}}{(1+R_{GMV,i})^2+(1+s_i)V_{GMV,i}}\,.\]
The last expression is obtained by applying the Sherman-Morrison formula. At last, we recall $\bQ_{T-t+1}\bii=\bzero$ and get (\ref{weights_T-t_cor}). Thus the corollary is proved.
\vspace{0.3cm}

\noindent \textbf{Proof of Theorem 2:}

\noindent The expression of the optimal weights at period $T-1$ is given in (\ref{weights_T-1_Sec3}). The rest of the theorem's statement is proved by using the mathematical induction on the expressions of the portfolio weights and the value function. We use similar notations as in the proof of Theorem 1. Let $\brA_i=E_{i-1}[(1-\tilde{s}_{i+1})\brx_{i}\brx^\prime_{i}]$ for $i=1,\ldots,T-1$ and $\brA_T=\bSigma_T-\brm_T\brm_T^\prime,$
\begin{equation}
\brm^*_i=\left\{
  \begin{array}{l l}
    \brm_{T} &\quad\text{for}~~~~i=T\\
    E_{i-1}[(1-\tilde{s}_{i+1})\tbx_{i}]& \quad \text{for}~~~ i=1,\ldots,T-1\,,
  \end{array} \right.
\end{equation}
and $\breve{s}_i=\brm_i^{*\;\prime} \brA^{-1}_i\brm^*_i$ for $i=2,...,T$.

Let $\bw^{*\;\prime}_{T-1}$ be the optimal portfolio weight calculated at period $T-1$ in the case of a riskless asset as expressed in (\ref{weights_T-1_Sec3}).
First, we calculate the value function at period $T-2$. It holds that
{\scriptsize
\begin{eqnarray*}
&&V(T-2,W_{T-2},\mathcal{F}_{T-2})\\
&=&\max\limits_{\bw_{T-2}}E_{T-2}\Big{[}W_{T-1}\left(R_{f,T}+\bw^{*\;\prime}_{T-1}\brm_{T}\right)-\frac{\alpha}{2}W^2_{T-1}\left(\bw^{*\;\prime}_{T-1}\brA_{T}\bw^*_{T-1}+R^2_{f,T}+2R_{f,T}\bw^{*\;\prime}_{T-1}\brm_{T}\right)\Big{]}\\
&=&\max\limits_{\bw_{T-2}}E_{T-2}\Big{[}W_{T-1}\left(R_{f,T}+\left(\frac{1}{\alpha W_{T-1}}-R_{f,T}\right)\brm^\prime_T\brA^{-1}_T\brm_{T}\right)\\
&-&\frac{\alpha}{2}W^2_{T-1}\left(\left(\frac{1}{\alpha W_{T-1}}-R_{f,T}\right)\brm^\prime_T\brA^{-1}_T\brA_{T}\left(\frac{1}{\alpha W_{T-1}}-R_{f,T}\right)\brA^{-1}_T\brm_T+R^2_{f,T}\right.\\
&+&\left.2R_{f,T}\left(\frac{1}{\alpha W_{T-1}}-R_{f,T}\right)\brm^\prime_T\brA^{-1}_T\brm_{T}\right)\Big{]}\,.
\end{eqnarray*}
}
Using the definition of $\breve{s}_T$ we obtain
{\footnotesize
\begin{eqnarray*}
&&V(T-2,W_{T-2},\mathcal{F}_{T-2})=\max\limits_{\bw_{T-2}}E_{T-2}\Big{[}W_{T-1}R_{f,T}(1-\breve{s}_T)+\frac{\breve{s}_T}{\alpha}\\
&-&\frac{\alpha}{2}W^2_{T-1}\left(\left(\frac{1}{\alpha W_{T-1}}-R_{f,T}\right)^2\breve{s}_T+R^2_{f,T}+2R_{f,T}\left(\frac{1}{\alpha W_{T-1}}-R_{f,T}\right)\breve{s}_T\right)\\
&=&\max\limits_{\bw_{T-2}}E_{T-2}\Big{[}W_{T-1}R_{f,T}(1-\breve{s}_T)+\frac{\breve{s}_T}{2\alpha}-\frac{\alpha}{2}W^2_{T-1}R^2_{f,T}\left(1-\breve{s}_T\right)\Big{]}\\
%&=&\max\limits_{\{\bw_{T-2}\}}\Big{[}W_{T-2}\left(r_{f,T}+\bw^{\prime}_{T-2}\brm_{T}\right)-\frac{\alpha}{2}W^2_{T-2}\left(\bw^{\prime}_{T-2}\bSigma_{T-1}\bw_{T-2}+(\bw^{\prime}_{T-2}\brm_{T})^2+r^2_{f,T}+2r_{f,T}\bw^{\prime}_{T-2}\brm_{T}\right)\Big{]}\\
&=&\max\limits_{\bw_{T-2}}\Big{[}W_{T-2}R_{f,T}\left(E_{T-2}[1-\breve{s}_T]R_{f,T-1}+\bw^{\prime}_{T-2}\brm^*_{T-1}\right)\\
&-&\frac{\alpha}{2}W^2_{T-2}R^2_{f,T}\left(\bw^{\prime}_{T-2}\bA_{T-1}\bw_{T-2}+E_{T-2}[1-\breve{s}_T]R^2_{f,T-1}+2R_{f,T-1}\bw^{\prime}_{T-2}\brm^*_{T-1}\right)+\frac{E_{T-2}[\breve{s}_T]}{2\alpha}\Big{]}\,.
\end{eqnarray*}
}

The last expression is similar to the value function at the period $T-1$. Hence, it is maximized on the weights $\bw_{T-2}^*$ expressed as
\begin{equation}\label{weights_T-2}
\bw^*_{T-2}=\left(\frac{1}{\alpha W_{T-2}}(R_{f,T})^{-1}-R_{f,T-1}\right)\brA^{-1}_{T-1}\brm^*_{T-1}\,.
\end{equation}

Hence, the basis of induction are the following expressions
\begin{eqnarray*}
&&V(T-2,W_{T-2},\mathcal{F}_{T-2})=\max\limits_{\{\bw_{T-2}\}}\Big{[}W_{T-2}R_{f,T}\left(b_TR_{f,T-1}+\bw^{\prime}_{T-2}\brm^*_{T-1}\right)\\
&-&\frac{\alpha}{2}W^2_{T-1}R^2_{f,T}\left(\bw^{\prime}_{T-2}\bA_{T-1}\bw_{T-2}+b_TR^2_{f,T-1}+2R_{f,T-1}\bw^{\prime}_{T-2}\brm^*_{T-1}\right)+F(\breve{s}_T)\Big{]}\\
&&\bw^*_{T-2}=\left(R_{f,T-1}-\frac{1}{\alpha W_{T-2}}(R_{f,T})^{-1}\right)\brA^{-1}_{T-1}\brm^*_{T-1}\\
\end{eqnarray*}
with $F(\breve{s}_T)=\frac{E_{T-2}[\breve{s}_T]}{2\alpha}$ and $b_T=E_{T-2}[1-\breve{s}_T]$.

In the induction hypothesis we assume that the statement holds for $t=n$, i.e.,
{\scriptsize
\begin{eqnarray*}
&&V(T-n,W_{T-n},\mathcal{F}_{T-n})=\max\limits_{\{\bw_{T-n}\}}\Big{[}W_{T-n}\left(\prod\limits_{i=T-n+2}^{T}R_{f,i}\right)\left(b_{T-n+2}R_{f,T-n+1}+\bw^{\prime}_{T-n}\brm^*_{T-n+1}\right)\\
&-&\frac{\alpha}{2}W^2_{T-n}\left(\prod\limits_{i=T-n+2}^{T}R^2_{f,i}\right)\left(\bw^{\prime}_{T-n}\bA_{T-n+1}\bw_{T-n}+b_{T-n+2}R^2_{f,T-n+1}+2R_{f,T-n+1}\bw^{\prime}_{T-n}\brm^*_{T-n+1}\right)\\
&+&F(\breve{s}_T,\ldots,\breve{s}_{T-n+2})\Big{]},\\
&&\bw^*_{T-n}=\left(\frac{1}{\alpha W_{T-n}}\left(\prod\limits_{i=T-n+2}^{T}R_{f,i}\right)^{-1}-R_{f,T-n+1}\right)\brA^{-1}_{T-n+1}\brm^*_{T-n+1}\,.
\end{eqnarray*}
}
with $F(\breve{s}_T,\ldots,\breve{s}_{T-n+2})=\frac{1}{2\alpha}\left(E_{T-2}[\breve{s}_T]+\sum\limits_{m=T-n+2}^{T-1}\prod\limits_{i=m}^{T-1}b_iE_{m-2}[\breve{s}_m]\right)$ and $b_i=E_{i-2}[1-\breve{s}_i]$.

In the inductive step we prove that the last identities also hold for $t=n+1$. It is sufficient to derive the value function for period $T-(n+1)$ which is given by
{\tiny
\begin{eqnarray*}
&&V(T-(n+1),W_{T-(n+1)},\mathcal{F}_{T-(n+1)})\\
&=&\max\limits_{\bw_{T-(n+1)}}E_{T-(n+1)}\Big{[}W_{T-n}\left(\prod\limits_{i=T-n+2}^{T}R_{f,i}\right)\left(b_{T-n+2}R_{f,T-n+1}+\bw^{*\;\prime}_{T-n}\brm^*_{T-n+1}\right)\\
&-&\frac{\alpha}{2}W^2_{T-n}\left(\prod\limits_{i=T-n+2}^{T}R^2_{f,i}\right)\left(\bw^{*\;\prime}_{T-n}\bA_{T-n+1}\bw^*_{T-n}+b_{T-n+2}R^2_{f,T-n+1}+2R_{f,T-n+1}\bw^{*\;\prime}_{T-n}\brm^*_{T-n+1}\right)\\
&-&F(\breve{s}_T,\ldots,\breve{s}_{T-n+2})\Big{]}\\
&=&\max\limits_{\bw_{T-(n+1)}}E_{T-(n+1)}\Big{[}W_{T-n}\prod\limits_{i=T-n+2}^{T}R_{f,i}\left(b_{T-n+2}R_{f,T-n+1}+\left(\frac{1}{\alpha W_{T-n}}\left(\prod\limits_{i=T-n+2}^{T}R_{f,i}\right)^{-1}-R_{f,T-n+1}\right)\right.\\
&\times&\left.\brm^{*\;\prime}_{T-n+1}\brA^{-1}_{T-n+1}\brm^*_{T-n+1}\right)\\
&-&\frac{\alpha}{2}W^2_{T-n}\prod\limits_{i=T-n+2}^{T}R^2_{f,i}\left(\left(\frac{1}{\alpha W_{T-n}}\left(\prod\limits_{i=T-n+2}^{T}R_{f,i}\right)^{-1}-R_{f,T-n+1}\right)\brm^{*\;\prime}_{T-n+1}\brA^{-1}_{T-n+1}\bA_{T-n+1}\right.\\
&\times&\left.\left(\frac{1}{\alpha W_{T-n}}\left(\prod\limits_{i=T-n+2}^{T}R_{f,i}\right)^{-1}-R_{f,T-n+1}\right)\brA^{-1}_{T-n+1}\brm^*_{T-n+1}+b_{T-n+2}R^2_{f,T-n+1}+\right.\\
&+&\left.2R_{f,T-n+1}\left(\frac{1}{\alpha W_{T-n}}\left(\prod\limits_{i=T-n+2}^{T}R_{f,i}\right)^{-1}-R_{f,T-n+1}\right)\brm^{*\;\prime}_{T-n+1}\brA^{-1}_{T-n+1}\brm^{*}_{T-n+1}\right)+F(\breve{s}_T,\ldots,\breve{s}_{T-n+2})\Big{]}\,.
\end{eqnarray*}
}

Using the definition of $\breve{s}_i$ and denoting $\xi=\prod\limits_{i=T-n+2}^{T}R_{f,i}$ we receive
{\scriptsize
\begin{eqnarray*}
&&V(T-(n+1),W_{T-(n+1)},\mathcal{F}_{T-(n+1)})\\
&=&\max\limits_{\{\bw_{T-(n+1)}\}}E_{T-(n+1)}\Big{[}W_{T-n}R_{f,T-n+1}\xi b_{T-n+2}(1-\breve{s}_{T-n+1})+\frac{b_{T-n+2}}{\alpha}\breve{s}_{T-n+1}\\
&+&F(\breve{s}_T,\ldots,\breve{s}_{T-n+2})-\frac{\alpha}{2}W^2_{T-n}b_{T-n+2}\xi^2\left(\left(\frac{\xi^{-1}}{\alpha W_{T-n}}-R_{f,T-n+1}\right)^2\breve{s}_{T-n+1}+R^2_{f,T-n+1}\right.\\
&+&\left.2R_{f,T-n+1}\left(\frac{\xi^{-1}}{\alpha W_{T-(n+1)}}-R_{f,T-n+1}\right)\breve{s}_{T-n+1}\right)\Big{]}\\
&=&\max\limits_{\{\bw_{T-(n+1)}\}}E_{T-(n+1)}\Big{[}W_{T-n}\xi R_{f,T-n+1} b_{T-n+2} (1-\tilde{s}_{T-n+1})\\
&+&\left(\frac{\breve{s}_{T-n+1}}{2\alpha}b_{T-n+2}+F(\breve{s}_T,\ldots,\breve{s}_{T-n+2})\right)-\frac{\alpha}{2}W^2_{T-n}(\xi R_{f,T-n+1})^{2}b_{T-n+2} (1-\tilde{s}_{T-n+1})\Big{]}\\
&=&\max\limits_{\{\bw_{T-(n+1)}\}}\Big{[}W_{T-(n+1)}\xi R_{f,T-n+1} b_{T-n+2}\left( E_{T-(n+1)}(1-\tilde{s}_{T-n+1})R_{f,T-n}+\bw^{\prime}_{T-(n+1)}\brm^*_{T-n}\right)\\
&-&\frac{\alpha}{2}W^2_{T-(n+1)}(\xi R_{f,T-n+1})^{2}b_{T-n+2} \left(\bw^{\prime}_{T-(n+1)}\bA_{T-n}\bw_{T-(n+1)}+ E_{T-(n+1)}(1-\tilde{s}_{T-n+1})R^2_{f,T-n}\right.\\
&+&\left.2R_{f,T-n}\bw^{\prime}_{T-(n+1)}\brm^*_{T-n}+F(\breve{s}_T,\ldots,\breve{s}_{T-n+1})\right)\,.
\end{eqnarray*}
}
where $F(\breve{s}_T,\ldots,\breve{s}_{T-n+1})=F(\breve{s}_T,\ldots,\breve{s}_{T-n+2})+\frac{1}{2}\frac{E_{T-(n+1)}[\breve{s}_{T-n+1}]}{\alpha}b_{T-n+2}$.

It is a desired form of the value function at period $T-(n+1)$. Because this expression is similar to the value function at period $T-n$, we get the following formula for the weights at period $T-(n+1)$
\begin{equation*}
\bw^*_{T-(n+1)}=\left(\frac{(\xi R_{f,T-n+1})^{-1}}{\alpha W_{T-n}}-R_{f,T-n}\right)\brA^{-1}_{T-n}\brm^*_{T-n}
\,,
\end{equation*}

Substituting $\xi=\prod\limits_{i=T-n+2}^{T}R_{f,i}$ leads to the expression given in the statement of Theorem 2. The theorem is proved.
\\[0.3cm]

\begin{proposition}
Let $\bx$ be a random vector with mean $\bmu$ and positive definite covariance matrix $\bSigma$. Let $\brA=\bSigma+\brm\brm^\prime$ and $\brm=\bmu-r_{f}\bii$. If
\begin{equation} \label{wrA}
\bw=\tilde{\gamma}^{-1}\brA^{-1}\brm
\end{equation}
then
\begin{equation} \label{wrE}
\bw=\gamma^{-1}\bSigma^{-1}\brm ~~~~\text{with}~~~~\gamma^{-1}=\frac{\tilde{\gamma}^{-1}}{1+\brm^\prime\bSigma^{-1}\brm}\,.
\end{equation}

\end{proposition}

\noindent \textbf{Proof of Proposition 2:}

\noindent The application of the Sherman-Morrison formula, i.e.,
$$\brA^{-1}=(\bSigma+\brm\brm^{\prime})^{-1}=\bSigma^{-1}-\frac{\bSigma^{-1}\brm\brm^{\prime}\bSigma^{-1}}{1+\brm^\prime\bSigma^{-1}\brm}$$
leads to
\begin{eqnarray*}
\bw&=&\tilde{\gamma}^{-1}\bSigma^{-1}\brm-\tilde{\gamma}^{-1}\frac{\bSigma^{-1}\brm\brm^{\prime}\bSigma^{-1}}{1+\brm^\prime\bSigma^{-1}\brm}\brm=\frac{\tilde{\gamma}^{-1}}{1+\brm^\prime\bSigma^{-1}\brm}\bSigma^{-1}\brm\,,
\end{eqnarray*}
what completes the proof of the proposition.\\[0.3cm]

\noindent \textbf{Proof of Corollary 2:}

\noindent Under the assumption of independence
\begin{equation}\label{brevA}
\brA_{T-t+1}= \left\{
  \begin{array}{l l}
    \bSigma_{T}+\brm_{T}\brm_{T}^\prime &\quad\text{for}~~~~t=1\\
    (1-\tilde{s}_{T-t+2}) (\bSigma_{T-t+1}+\brm_{T-t+1}\brm_{T-t+1}^\prime ) & \quad \text{for}~~~ t=2,\ldots,T\\
  \end{array} \right.
\end{equation}
and
\begin{equation}\label{brevmu}
\brm_{T-t+1}^*= \left\{
  \begin{array}{l l}
    \brm_{T} &\quad\text{for}~~~~t=1\\
    (1-\tilde{s}_{T-t+2})\brm_{T-t+1} & \quad \text{for}~~~ t=2,\ldots,T\\
  \end{array} \right.\,.
\end{equation}

Then the statement of the corollary follows from Proposition 2 if $\bSigma$ is replaced by $\bSigma_{T-t+1}$ and $\brm$ by $\brm_{T-t+1}$, and
\[\tilde{\gamma}^{-1}=\Big{[}\frac{1}{\alpha W_{T-t}}\left(\prod\limits_{i=T-t+2}^{T}R_{f,i}\right)^{-1}-R_{f,T-t+1}\Big{]}\,.\]
\vspace{0.3cm}

\noindent \textbf{Proof of Theorem 3:}

\noindent The results of Theorem 3 follow Theorem 2 and the application of the Sherman-Morrison formula.\\

\end{document}